\documentclass[trackchanges]{aastex701}
\usepackage{amsmath}

\begin{document}

\title{Radial Evolution of Near-Sun Magnetic Switchbacks \text{Alfv\'enicity}, Occurrence Rate, and Size}

\author[orcid=0000-0002-7685-1528]{Xiaolei Li}
\affiliation{Physics Department, Auburn University, Auburn, AL, 36832, USA}
\email[show]{xzl0125@auburn.edu}  

\author[orcid=0000-0002-2582-7085]{Chen Shi} 
\affiliation{Physics Department, Auburn University, Auburn, AL, 36832, USA}
\email[show]{chenshi@auburn.edu}

\author[0009-0009-0162-2067]{Yuliang Ding}
\affiliation{Department of Earth, Planetary, and Space Sciences, University of California, Los Angeles, Los Angeles, CA 90095, USA}
\email[show]{dingyl@g.ucla.edu}

\begin{abstract}
Magnetic switchbacks, characterized by reversals of magnetic field direction, are widely observed in the inner heliosphere by Parker Solar Probe (PSP). With PSP reaching perihelia near 10$R_\odot$, observations from the first 24 encounters enable studies of near-Sun switchback evolution at $r>10R_\odot$. We construct a switchback catalog within $10<r<55R_\odot$ by identifying magnetic field reversals with stable field magnitude and strahl-electron polarity. Statistical analysis shows that switchback Alfvénicity decreases with increasing radial distance, consistent with solar wind evolution beyond the Alfvén critical point. Meanwhile, switchback occurrence rate and spatial size increase with distance, suggesting continued generation and expansion during solar wind propagation. At a given radial distance, the fraction of solar wind containing switchbacks is positively correlated with background solar wind radial velocity ($V_R$) and Alfvén Mach number ($M_A$), while the local occurrence rate is mainly controlled by $M_A$. These results suggest that switchback patches preferentially form in faster and higher-$M_A$ solar wind. The spatial size of switchbacks shows no clear dependence on $M_A$ or $V_R$, implying that their size evolution is probably not determined by source conditions. Solar activity influences switchback evolution through changes in background solar wind properties, with a larger fraction of higher-$M_A$ switchbacks during solar minimum. We further identify anisotropy relative to the background magnetic field direction: the local occurrence rate and spatial size are approximately 1.5 times as large in the perpendicular direction as in the parallel direction, indicating distinct magnetic topology of switchback patches.
\end{abstract}



\section{Introduction} \label{sec:intro}

The Parker Solar Probe (PSP) mission has, for the first time, reached the frontier of the near-Sun heliosphere, sampling solar wind plasma at heliocentric distances below 0.3 AU. One of the most prominent solar wind phenomena revealed by PSP turbulence observations is the presence of switchbacks, which are defined as localized reversals of the magnetic field polarity \citep[e.g.,][]{bale2019highly,kasper2019alfvenic}. Several characteristics observed during switchbacks, including the nearly constant magnetic field magnitude and the strong correlation between velocity and magnetic field fluctuations, suggest that switchbacks are highly Alfvénic structures \citep{kasper2019alfvenic,larosa2021switchbacks}, resembling outward-propagating Alfvén waves originating from the Sun. The strong Alfvénic turbulence embedded within switchbacks may provide additional energy to the turbulent cascade \citep{de2020switchbacks,hernandez2021impact} and may contain sufficient energy to contribute to solar wind heating and acceleration, serving as an important energy source in the heliosphere \citep{halekas2023quantifying,rivera2024situ}.


Many theories have been proposed to explain the formation mechanisms of near-Sun switchbacks \citep{wyper2026magnetic}. One class of theories suggests that switchbacks are generated during solar wind expansion as the result of the evolution of Alfvénic fluctuations originating from the lower corona. These fluctuations may be excited by convective motions \citep{tziotziou2023vortex}, reconnection between open and closed magnetic field structures \citep{zank2020origin}, or solar jets \citep{touresse2024propagation}. Alternatively, other theoretical models propose that velocity shears in the inner heliosphere can distort magnetic field lines, producing local field reversals and forming switchbacks on both large scales \citep{schwadron2021switchbacks,toth2023theory} and small scales \citep{ruffolo2020shear}. Further observational evidence is required to distinguish between these competing formation scenarios.


Recent studies have investigated the properties of near-Sun switchbacks at heliocentric distances of $r<0.3$ AU, including their Alfvénicity, occurrence rate, and characteristic scales \citep{badman2026properties,mallet2026evolution}. The normalized cross helicity $\sigma_c$ and residual energy $\sigma_r$, both ranging from -1 to 1, are commonly used to quantify the Alfvénicity of switchback-associated turbulence. For a simple Alfvén wave, $\sigma_c$ approaches $\pm1$ and $\sigma_r$ approaches 0, indicating a high degree of Alfvénicity. Parker Solar Probe observations show that intervals containing switchbacks generally exhibit relatively high Alfvénicity, characterized by $\sigma_c$ slightly lower than 1 and weakly negative $\sigma_r$ values \citep{wu2021large,agapitov2023constraints}, with higher Alfvénicity at switchback center than boundary wind \citep{agapitov2023constraints, 2022ApJ...934..152S}. This suggests enhanced nonlinear interactions between counter-propagating Alfvénic fluctuations within switchback structures.

Switchbacks are not randomly distributed in the solar wind, but are modulated by the spacecraft location, temporal evolution, and properties of the background solar wind. They are frequently observed in spatially and temporally clustered patches \citep{horbury2020solar,fargette2021characteristic}, where a series of magnetic deflections can persist for hours due to large-scale modulation effects \citep{2021ApJ...923..174B,2022ApJ...934..152S}, resulting in locally enhanced occurrence rates. Statistical studies have shown that the occurrence rate of switchbacks is correlated with solar wind speed \citep{mozer2021origin,jagarlamudi2023occurrence}, Alfvén Mach number \citep{payne2026evolution}, and heliocentric distance \citep{pecora2022magnetic}, modulated by the fluctuation scale \citep{tenerani2021evolution}.

Regarding switchback sizes, statistical analyses indicate that their duration distributions exhibit power-law tails with slopes approaching -2 \citep{pecora2022magnetic}, and the distributions become flatter with increasing heliocentric distance \citep{tenerani2021evolution}. Near the Sun, switchback spatial scales span approximately $10^3$--$10^7$ km, with a median size of about $10^5$ km ($\sim0.1R_\odot$) \citep{meng2022analysis}. Their characteristic size has also been found to increase with heliocentric distance \citep{mozer2021origin}.

Since the beginning of the Parker Solar Probe (PSP) mission in 2018, several catalogs of near-Sun switchbacks have been constructed from more than four PSP encounters by identifying magnetic field deflections exceeding 90$^\circ$. \cite{mozer2021origin} identified switchbacks during PSP encounters 3--7 at heliocentric distances of $20<r<100R_\odot$ by selecting events with magnetic field rotations exceeding 90$^\circ$ relative to the Parker spiral direction and applying a cutoff angle of 60$^\circ$. \cite{tenerani2021evolution} defined the background magnetic field direction as the temporal average within a sliding window ranging from 30 min to 12 hr, and identified switchbacks in the first six PSP encounters at $r>0.1$ AU as intervals with magnetic deflections larger than 90$^\circ$ and low magnetic compressibility ($\delta \boldsymbol{B}/|\boldsymbol{B}|<0.2$). \cite{pecora2022magnetic} used magnetic field measurements at 10 s and 60 s resolutions and adopted a similar background-field determination method with sliding windows of 3 hr and 6 hr to identify switchbacks during PSP encounters 1--8 at $r>40R_\odot$. \cite{huang2023structure} further considered the polarity of suprathermal electron pitch-angle distributions to exclude current sheet crossings and constructed a catalog containing 1748 switchbacks from the first eight PSP encounters.

With the decrease of the PSP perihelion distance to below 20$R_\odot$ after encounter 8 in 2021 and to approximately 10$R_\odot$ after encounter 17 in 2023, together with the observations of sub-Alfvénic solar wind, new switchback catalogs are required to investigate switchback properties much closer to the Alfvén critical point \citep{sioulas2025propagation,sioulas2026generation}, which separates the sub-Alfvénic and super-Alfvénic solar wind regimes. Such catalogs are also essential for studying switchback evolution under different solar activity conditions, from solar minimum to solar maximum.

In this work, we use data from the first 24 PSP encounters to construct a comprehensive near-Sun switchback catalog and investigate the radial evolution of switchback properties and their impacts on the surrounding solar wind. In Section \ref{sec:D&M}, we describe the data sets and methodology used to identify switchbacks and evaluate the selected parameters. The resulting switchback catalog is presented in Section \ref{sec:result}. The radial evolution of switchback properties is discussed in Section \ref{ssec:rev}, while the associated impacts, including the Alfvén Mach number, solar wind velocity, solar activity, and magnetic anisotropy, on the switchback properties are presented in Sections \ref{ssec:rMA}--\ref{ssec:ralpha}. Finally, Section \ref{sec:D&C} provides the discussion and conclusions.

\section{Data and Method} \label{sec:D&M}
\subsection{Data selection} \label{ssec:D}

We focus on PSP measurements obtained near perihelia during encounters 1--24 (E01--E24). For the magnetic field vector $\boldsymbol{B}$, we use level-2 fluxgate magnetometer measurements from the FIELDS instrument, downsampled to four samples per cycle with a sampling rate of approximately 4.58 samples s$^{-1}$ \citep{2016SSRv..204...49B,fox2016solar}.

For plasma density measurements, we use level-3 electron density ($n_e$) derived from quasi-thermal noise spectroscopy analysis of the Radio Frequency Receivers (RFS) data onboard PSP. The cadence of the density measurements is 7 s during E01--E05 and 3.5 s during E06--E24, except for E08 and E11 \citep{2017JGRA..122.2836P,Kruparova_2023}. Assuming an alpha-to-proton number density ratio of $A_{He}=n_\alpha/n_p\times100\%=4\%$ like \cite{huang2023parker}, the proton number density is estimated to be $n_p=n_e/(1+2A_{He})$. 

For the solar wind velocity $\boldsymbol{V}$, we use proton velocity measurements from the level-3 data of the Solar Probe ANalyzer for Ions (SPAN-I). We note that the transverse solar wind velocity component $V_T$ measured by SPAN-I can be biased due to instrumental field-of-view limitations \citep{livi2022solar}; therefore, we neglect the $V_T$ component in the following analysis. This does not alter the main results presented in this work, as confirmed by repeating the analysis with $V_T$ included. We also use level-3 electron pitch-angle distribution measurements from the Solar Probe ANalyzer for Electrons (SPAN-E), with a time cadence as high as 3.5 s \citep{fox2016solar,2016SSRv..204..131K}.

\begin{figure*}[ht!]
\plotone{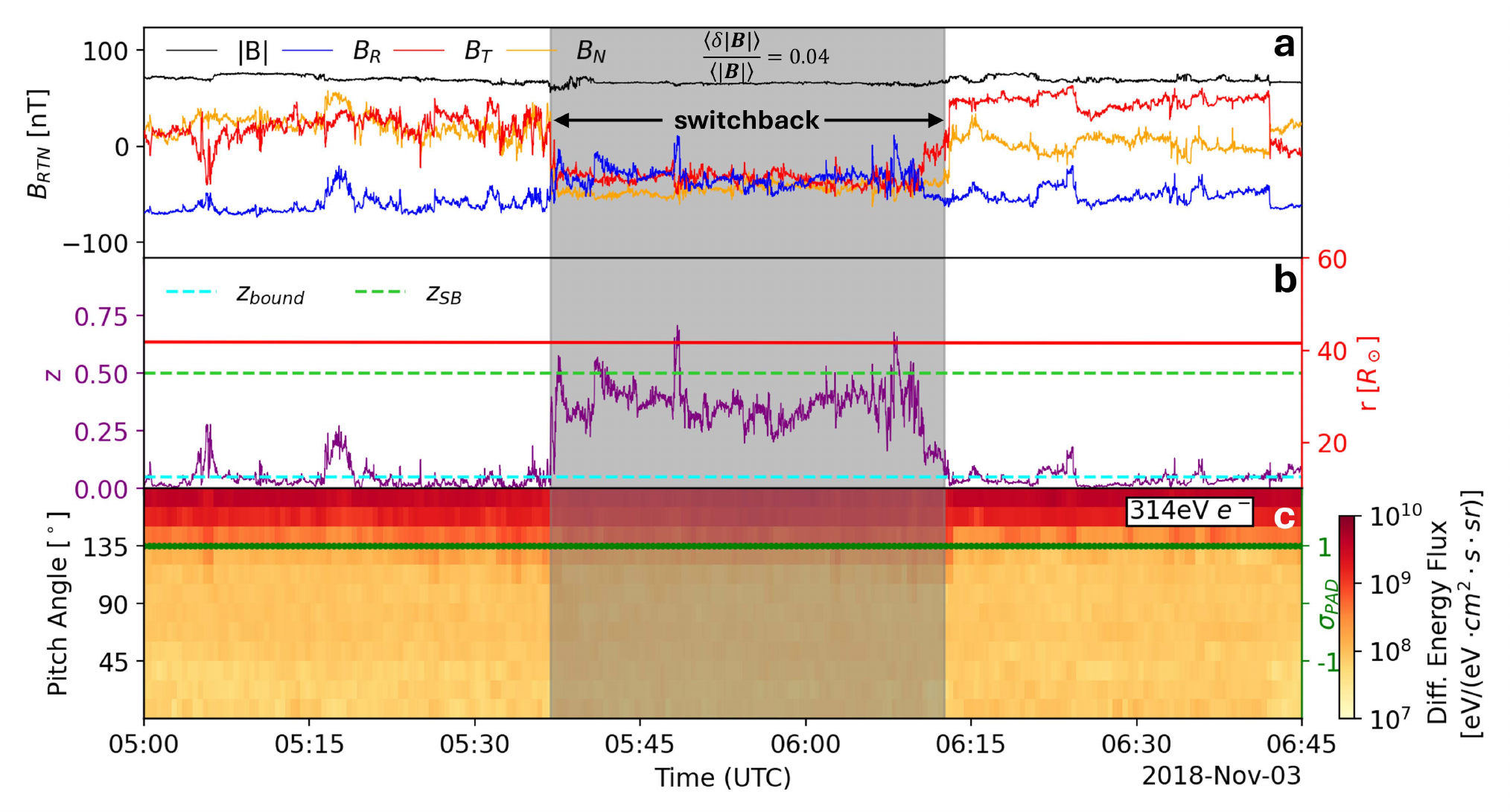}
\caption{Example of an identified switchback shown as the gray-shaded region. (a) Magnetic field magnitude $|\boldsymbol{B}|$ (black) and the three magnetic field components $B_R$ (blue), $B_T$ (red), and $B_N$ (orange). (b) The deflection parameter $z$ (purple), and the heliocentric distance $r$ (red). $z_{bound}=0.05$ (cyan) and $z_{SB}=0.5$ (lime) are identification threshold and boundary threshold. (c) The 314 eV electron differential energy flux pitch-angle distribution. The green curve represents $\sigma_{PAD}$ as defined in Section \ref{ssec:Me}. This event satisfies all three identification criteria described in Section \ref{ssec:Me}: a large magnetic field deflection with local $z>z_{SB}$ and boundary $z=z_{bound}$, relatively constant magnetic field magnitude with $\langle\delta|\boldsymbol{B}|\rangle/\langle|\boldsymbol{B}|\rangle=0.04<0.15$, and a stable strahl electron pitch-angle direction with $\sigma_{PAD}=1$.
\label{fig:1}}
\end{figure*}

\subsection{Switchbacks identification method} \label{ssec:Me}

A switchback is identified based on the following three criteria, as illustrated in Figure \ref{fig:1}.

(1) Large magnetic field deflection. A switchback must contain a magnetic field deflection exceeding $90^\circ$. To identify regions with large magnetic deflections, we first downsample the magnetic field data to a cadence of $\Delta t=1$ s and define the normalized magnetic field deflection parameter $z$ as \citep{de2020switchbacks}
\begin{equation}
z=\frac{1}{2}(1-\cos{\theta}),
\quad
\theta=\arccos{\frac{\boldsymbol{B}\cdot\boldsymbol{B}_0}{|\boldsymbol{B}||\boldsymbol{B}_0|}},
\label{eq1}
\end{equation}
where $\theta$ is the angle between the local magnetic field $\boldsymbol{B}$ and the background magnetic field $\boldsymbol{B}_0$. A value of $z\geq0.5$ corresponds to a magnetic field deflection larger than $90^\circ$. Therefore, intervals with local $z>z_{SB}=0.5$ are selected as potential switchbacks. To determine the boundaries of each switchback, we apply a threshold $z_{bound}=0.05$ to distinguish switchback intervals from the background solar wind with weak magnetic deflections ($z<0.05$) \citep{de2020switchbacks}.

To obtain a more accurate estimate of $\boldsymbol{B}_0$ and $z$, we initially calculate $\boldsymbol{B}_0$ as the median magnetic field vector within a running window of $T_0=6$ hr, following \cite{de2020switchbacks} and \cite{pecora2022magnetic}. We then calculate $z$ using Equation \ref{eq1}. The background magnetic field $\boldsymbol{B}_0$ is subsequently updated by interpolating the magnetic field vectors at background solar wind intervals with $z<0.05$. Based on the updated $\boldsymbol{B}_0$, $z$ value is recalculated. The last two steps are iterated three times to obtain the final $\boldsymbol{B}_0$ and $z$ used for switchback identification.

(2) Low magnetic field magnitude variation. The magnetic field magnitude $|\boldsymbol{B}|$ is expected to remain approximately constant within a switchback \citep{kasper2019alfvenic,bale2019highly,tenerani2021evolution}. This criterion helps distinguish genuine switchbacks from current sheet crossings associated with large magnetic field magnitude variations. We quantify the relative perturbation of $|\boldsymbol{B}|$ as
$\frac{\langle\delta|\boldsymbol{B}|\rangle}{\langle|\boldsymbol{B}|\rangle}$,
where $\langle\delta|\boldsymbol{B}|\rangle$ and $\langle|\boldsymbol{B}|\rangle$ represent the standard deviation and mean value of $|\boldsymbol{B}|$ within a potential switchback, respectively. Potential switchbacks satisfying$
\frac{\langle\delta|\boldsymbol{B}|\rangle}{\langle|\boldsymbol{B}|\rangle}<0.15$
are retained.

(3) Stable strahl electron pitch-angle direction. The strahl electron pitch-angle direction is expected to remain unchanged throughout most switchbacks \citep[e.g.][]{kasper2019alfvenic}. Therefore, we use electron pitch-angle distributions (PADs) to exclude structures associated with magnetic flux ropes, which may exhibit bidirectional electron beams, and current sheets, which may show a reversal of the strahl direction across the structure \citep[e.g.][]{phan2020parker}.

We integrate the strahl electron PAD at an energy of 314 eV over pitch angles larger than $90^\circ$ and smaller than $90^\circ$, denoted as $n_e^>$ and $n_e^<$, respectively. The strahl electron streaming anisotropy and its polarity are defined as
\begin{equation}
   \widetilde{\sigma}_{PAD}=\frac{n_e^>-n_e^<}{n_e^>+n_e^<},\ 
    \sigma_{PAD}=
    \left\{
    \begin{array}{cc}
    1, & \widetilde{\sigma}_{PAD}>0.2,\\
    0, & |\widetilde{\sigma}_{PAD}|\le 0.2,\\
    -1, & \widetilde{\sigma}_{PAD}<-0.2.
    \end{array}
    \right.
\end{equation}

A potential switchback is confirmed if $\sigma_{PAD}$ maintains the same polarity for more than 80 percentage of the interval from $t_{b,SB}-\delta t$ to $t_{e,SB}+\delta t$, where $t_{b,SB}$ and $t_{e,SB}$ denote the start and end times of the potential switchback, respectively, and $\delta t=5$ min.

\subsection{Evaluation of parameters in identification method} \label{ssec:PS}
In this subsection, we evaluate the robustness of the parameters adopted in the switchback identification method described in Section \ref{ssec:Me}. The parameters examined include the magnetic field downsampling cadence $\Delta t$, the running-window duration used to determine the median background magnetic field $T_0$, the switchback identification threshold $z_{SB}$, the switchback boundary threshold $z_{bound}$, the threshold for relative magnetic field magnitude perturbation $\langle\delta|\boldsymbol{B}|\rangle/\langle|\boldsymbol{B}|\rangle$, and the threshold for the temporal percentage of dominant polarity of $\sigma_{PAD}$, denoted as $P_{\sigma_{PAD}}$.

\begin{figure*}[ht!]
\plotone{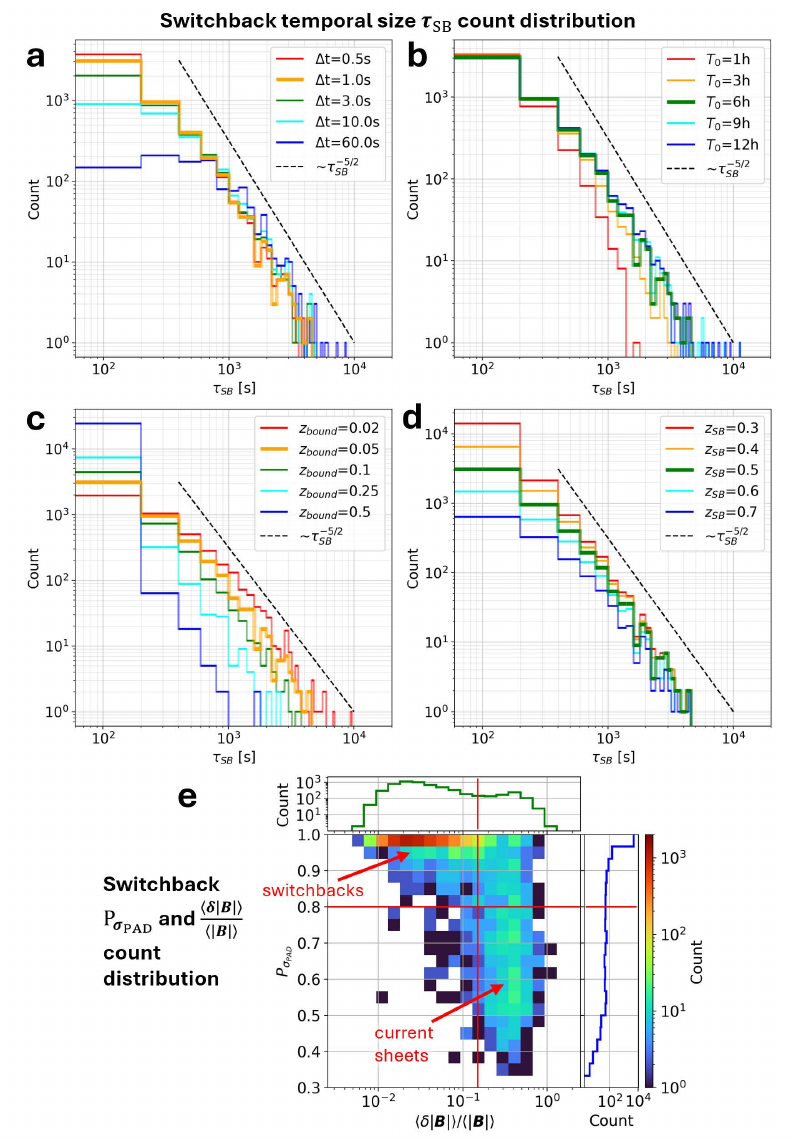}
\caption{Influence of identification parameters on the resulting switchback duration distribution. (a)--(d) The switchback duration distributions $\tau_{SB}$ for different values of $\Delta t$, $T_0$, $z_{bound}$, and $z_{SB}$, respectively. The thick curves correspond to the adopted parameter set used in this study. (e) The joint distribution of $\langle\delta|\boldsymbol{B}|\rangle/\langle|\boldsymbol{B}|\rangle$ and $P_{\sigma_{PAD}}$ for potential switchbacks identified using only criterion (1) in Section \ref{ssec:Me}, with corresponding one-dimensional distributions shown on the top (green) and right (blue) axes. The red vertical and horizontal lines mark the thresholds $\langle\delta|\boldsymbol{B}|\rangle/\langle|\boldsymbol{B}|\rangle=0.15$ and $P_{\sigma_{PAD}}=0.8$, dividing the parameter space into four populations. High $\langle\delta|\boldsymbol{B}|\rangle/\langle|\boldsymbol{B}|\rangle$ and low $P_{\sigma_{PAD}}$ correspond to likely current sheet crossings, while low $\langle\delta|\boldsymbol{B}|\rangle/\langle|\boldsymbol{B}|\rangle$ and high $P_{\sigma_{PAD}}$ correspond to switchbacks included in our catalog.
\label{fig:2}}
\end{figure*}

For the parameters $\Delta t$, $T_0$, $z_{bound}$, and $z_{SB}$, we vary their values and examine the corresponding changes in the switchback duration distribution $\tau_{SB}$, as shown in Figure \ref{fig:2}(a)--(d). The thick lines represent the distributions obtained with the adopted parameters. The resulting distribution follows approximately $N\sim\tau_{SB}^{-5/2}$, as indicated by the dashed reference line. The spectral slope of -2.5 obtained in this work for $0.05<r<0.25$ AU is steeper than the slope of approximately -2 reported by \cite{pecora2022magnetic} for $0.1<r<0.7$ AU, possibly because the spectral slope becomes flatter with increasing heliocentric distance \citep{tenerani2021evolution}.

As shown in Figure \ref{fig:2}a, the distributions for $\tau_{SB}>10^3$ s are nearly identical for $\Delta t$ values ranging from 0.5 s to 60 s. However, at $\tau_{SB}<10^3$ s, the number of identified switchbacks decreases as $\Delta t$ increases, resulting in an underestimation of short-duration switchbacks, especially for $\Delta t>3$ s. For $T_0>3$ hr, the duration distributions remain nearly unchanged, whereas shorter running windows ($T_0<3$ hr) lead to fewer identified switchbacks, as shown in Figure \ref{fig:2}b. In Figure \ref{fig:2}c, the number of switchbacks with $\tau_{SB}>200$ s decreases as $z_{bound}$ increases, while the opposite trend appears for $\tau_{SB}<200$ s. This results in abnormal spectral slopes when $z_{bound}>0.1$ or $z_{bound}<0.05$. As shown in Figure \ref{fig:2}d, both the identified switchback number and the absolute spectral index slightly decrease with decreasing $z_{SB}$, while $0.3<z_{SB}<0.7$ produces relatively stable distributions. Therefore, based on the sensitivity analysis of the $\tau_{SB}$ distribution, the adopted parameters ($\Delta t=1$ s, $T_0=6$ hr, $z_{bound}=0.05$, and $z_{SB}=0.5$) provide robust switchback identification.

To evaluate the thresholds for $\langle\delta|\boldsymbol{B}|\rangle/\langle|\boldsymbol{B}|\rangle$ and $P_{\sigma_{PAD}}$, we examine their one-dimensional distributions and joint two-dimensional distribution for potential switchbacks identified only by criterion (1), as shown in Figure \ref{fig:2}e. The vertical and horizontal red lines indicate the adopted thresholds of $\langle\delta|\boldsymbol{B}|\rangle/\langle|\boldsymbol{B}|\rangle=0.15$ and $P_{\sigma_{PAD}}=0.8$, respectively. The distribution of $\langle\delta|\boldsymbol{B}|\rangle/\langle|\boldsymbol{B}|\rangle$ (green curve) exhibits two peaks, with the selected threshold of 0.15 located near the minimum between the two peaks. The distribution of $P_{\sigma_{PAD}}$ consists of a boundary-enhanced component with a peak near 1 and a Gaussian-like component centered around 0.6, which are approximately separated by the adopted threshold of 0.8. In the two-dimensional distribution, these thresholds separate two major populations of potential switchbacks with large magnetic field deflections. One population exhibits multiple strahl electron polarities ($P_{\sigma_{PAD}}<0.8$) and enhanced magnetic field magnitude variations ($\langle\delta|\boldsymbol{B}|\rangle/\langle|\boldsymbol{B}|\rangle>0.15$), which are likely associated with current sheet crossings. The other population shows a dominant strahl electron polarity ($P_{\sigma_{PAD}}>0.8$) and weak magnetic field magnitude variations ($\langle\delta|\boldsymbol{B}|\rangle/\langle|\boldsymbol{B}|\rangle<0.15$), corresponding to the switchbacks selected in our catalog.

\section{Results and analysis} \label{sec:result}

\begin{deluxetable*}{rllc}
\digitalasset
\tablewidth{0pt}
\tablecaption{Start, End Times of the PSP Dataset Analyzed, and Identified Switchbacks Count near the Perihelion in E01-E24 \label{tab:1}}
\tablehead{
\colhead{Enc.} & \colhead{Start(UTC)} & \colhead{End(UTC)} & \colhead{Switchbacks Count} 
}
\startdata
1 & 2018-10-31 11:56 & 2018-11-11 18:58 & 483\\
2 & 2019-03-30 07:08 & 2019-04-10 14:10 & 250\\
3 & 2019-08-27 02:19 & 2019-09-07 09:21 & 262\\
4 & 2020-01-23 14:00 & 2020-02-04 05:14 & 252\\
5 & 2020-06-01 12:46 & 2020-06-13 03:59 & 202\\
6 & 2020-09-21 18:43 & 2020-10-02 23:48 & 267\\
7 & 2021-01-12 03:06 & 2021-01-23 08:12 & 263\\
8 & 2021-04-24 00:09 & 2021-05-04 17:24 & 238\\
9 & 2021-08-04 10:33 & 2021-08-15 03:47 & 210\\
10 & 2021-11-16 04:19 & 2021-11-26 12:31 & 370\\
11 & 2022-02-20 11:29 & 2022-03-02 19:41 & 245\\
12 & 2022-05-27 18:43 & 2022-06-07 02:56 & 117\\
13 & 2022-09-01 01:57 & 2022-09-11 10:09 & 324\\
14 & 2022-12-06 09:09 & 2022-12-16 17:22 & 247\\
15 & 2023-03-12 16:24 & 2023-03-23 00:37 & 129\\
16 & 2023-06-16 23:39 & 2023-06-27 07:52 & 140\\
17 & 2023-09-22 22:47 & 2023-10-03 00:09 & 126\\
18 & 2023-12-24 00:14 & 2024-01-03 01:36 & 137\\
19 & 2024-03-25 01:40 & 2024-04-03 18:48 & 193\\
20 & 2024-06-25 03:06 & 2024-07-05 04:28 & 59\\
21 & 2024-09-25 04:33 & 2024-10-05 05:55 & 81\\
22 & 2024-12-19 14:19 & 2024-12-29 09:28 & 70\\
23 & 2025-03-18 01:08 & 2025-03-27 20:17 & 104\\
24 & 2025-06-14 11:57 & 2025-06-24 07:06 & 213\\
\enddata
\end{deluxetable*}

Based on the method described in Section \ref{ssec:Me}, 4982 switchbacks are identified from PSP encounters E01--E24 near perihelia at heliocentric distances $r<55R_\odot$ (see $switchback\_catalog.csv$ in the attachment). The start and end times, together with the number of switchbacks for each encounter, are listed in Table \ref{tab:1}. The number of identified switchbacks per encounter decreases from more than 200 during 2018–2019 near solar maximum to fewer than 100 in the latter half of 2024.

\begin{figure*}[ht!]
\plotone{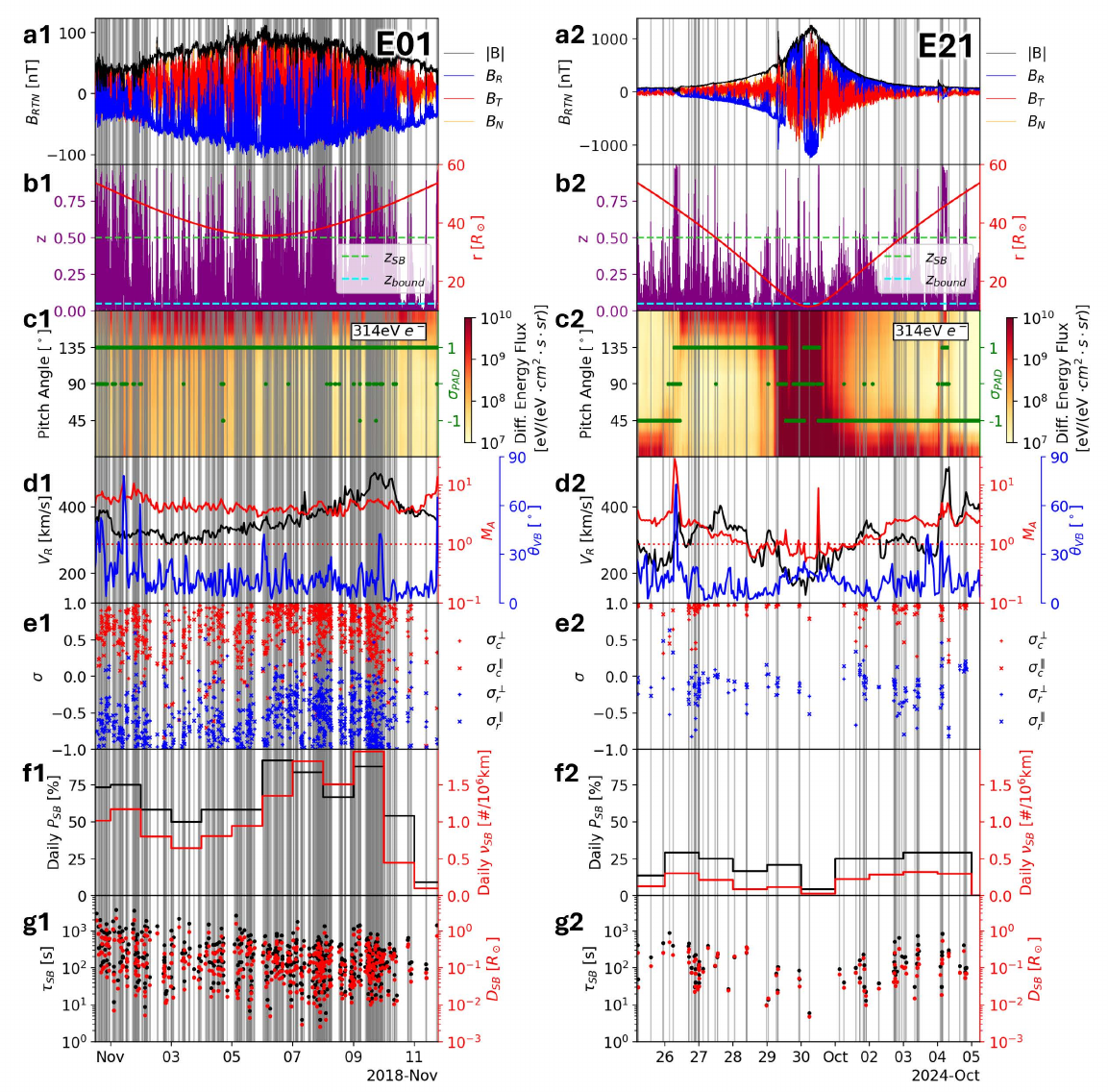}
\caption{Identified switchbacks (gray-shaded regions) in PSP encounter E01 during solar minimum (a1–g1) and E21 during solar maximum (a2–g2). (a)–(c) The same as in Figure \ref{fig:1}. (d) The radial solar wind velocity $V_R$ (black), the Alfvén Mach number $M_A$ (red), and the unsigned angle $\theta_{VB}$ (blue), defined as the angle between the PSP velocity and the magnetic field in the outward-propagating Alfvén-wave frame, for the background solar wind. The red dotted line indicates $M_A=1$. (e) The normalized perpendicular and parallel cross helicity and residual energy: $\sigma_c^\perp$ (red plus), $\sigma_c^\parallel$ (red cross), $\sigma_r^\perp$ (blue plus), and $\sigma_r^\parallel$ (blue cross). (f) The daily occurrence metrics: $P_{SB}$ (black), defined as the spatial fraction of ambient solar wind occupied by switchbacks, and $\nu_{SB}$ (red), the spatial occurrence rate of switchbacks in the ambient solar wind. (g) The switchback duration $\tau_{SB}$ (black) and spatial scale $D_{SB}$ (red) for each event.
\label{fig:3}}
\end{figure*}

Of all switchbacks in the catalog, 4450 with reliable solar wind density measurements from the PSP RFS are selected for further analysis of their properties and background solar wind conditions. Figures \ref{fig:3}a and \ref{fig:3}b show all identified and analyzed switchbacks during E01 (solar maximum) and E21 (solar minimum), respectively. The background solar wind conditions and switchback properties are presented in Figure \ref{fig:3}d-\ref{fig:3}g. The definitions of the relevant parameters are given below. For a generic parameter $X$, the background component $X_0$ is defined as the median value of $X$ outside switchbacks within a running window between 1/2 hr before and 1/2 hr after, while $\langle X\rangle$ denotes the average value of $X$ within a given switchback interval.

(1) Background solar wind quantities.

$V_R$: radial component of the background solar wind velocity, defined as $\boldsymbol{V}_{SW}=\boldsymbol{V}_0$.

$M_A$: Alfvén Mach number of the background solar wind, defined as $M_A=V_R/|\boldsymbol{V}_A|$, where the Alfvén velocity is given by
$
\boldsymbol{V}_A=\mathrm{sign}(B_{R0})\frac{\boldsymbol{B}_0}{\sqrt{\mu_0 (1+4A_{He}) n_{p0} m_H}},
$
representing the outward-propagating Alfvén speed in the background solar wind. Here $A_{He}=4\%$ is alpha-proton number density ratio as introduced in Section \ref{ssec:D}. 

$\theta_{VB}$: unsigned angle between $\boldsymbol{V}_{cross}$ and the background magnetic field $\boldsymbol{B}_0$, evaluated in the frame moving with the outward-propagating Alfvén speed:$
\theta_{VB}=\arccos\left(\frac{|\boldsymbol{V}_{cross}\cdot\boldsymbol{B}_0|}{|\boldsymbol{V}_{cross}||\boldsymbol{B}_0|}\right)$.
Here $\boldsymbol{V}_{cross}=\boldsymbol{V}_{PSP}-\boldsymbol{V}_{SW}-\boldsymbol{V}_A$, under the assumption that switchbacks correspond to outward-propagating Alfvénic fluctuations. This quantity characterizes the angle between the PSP crossing trajectory and the background magnetic field in the switchback reference frame (see Figure \ref{fig:8}m).

In the following statistical analysis, we use the median value of $V_R$, $M_A$, or $\theta_{VB}$ in each 1-hour intervals or in a switchback duration time as corresponding samples of background solar wind quantities.

(2) Switchback Alfvénicity properties.

$\sigma_c^{\perp}$, $\sigma_c^{\parallel}$, $\sigma_r^{\perp}$, and $\sigma_r^{\parallel}$: rolling-averaged normalized cross helicity and residual energy for fluctuations perpendicular and parallel to the background magnetic field \citep{sioulas2025propagation}. They are defined as
\begin{equation}
\begin{aligned}
\sigma_c^\perp &= \frac{2\langle \boldsymbol{v}_\perp \cdot \boldsymbol{b}_\perp \rangle}{\langle |\boldsymbol{v}_\perp|^2 \rangle + \langle |\boldsymbol{b}_\perp|^2 \rangle}, \quad
\sigma_c^\parallel = \frac{2\langle \boldsymbol{v}_\parallel \cdot \boldsymbol{b}_\parallel \rangle}{\langle |\boldsymbol{v}_\parallel|^2 \rangle + \langle |\boldsymbol{b}_\parallel|^2 \rangle}, \
\sigma_r^\perp &= \frac{\langle |\boldsymbol{v}_\perp|^2 \rangle - \langle |\boldsymbol{b}_\perp|^2 \rangle}{\langle |\boldsymbol{v}_\perp|^2 \rangle + \langle |\boldsymbol{b}_\perp|^2 \rangle}, \quad
\sigma_r^\parallel = \frac{\langle |\boldsymbol{v}_\parallel|^2 \rangle - \langle |\boldsymbol{b}_\parallel|^2 \rangle}{\langle |\boldsymbol{v}_\parallel|^2 \rangle + \langle |\boldsymbol{b}_\parallel|^2 \rangle}.
\end{aligned}
\end{equation}

Here, $\boldsymbol{v}=\boldsymbol{V}-\boldsymbol{V}_0$ and $\boldsymbol{b}=-\mathrm{sign}(B_{R0})(\boldsymbol{B}-\boldsymbol{B}_0)/\sqrt{\mu_0 (1+4A_{He})n_{p0} m_H}$. The subscripts $\perp$ and $\parallel$ denote components perpendicular and parallel to the background magnetic field $\boldsymbol{B}_0$, respectively. This sign convention ensures that $\boldsymbol{z}^\pm=\boldsymbol{v}\pm\boldsymbol{b}$ correspond to anti-sunward and sunward propagating fluctuations. Due to the known bias in the transverse velocity component $V_T$, only the radial and normal components are used in the calculation of $\boldsymbol{v}$ and $\boldsymbol{b}$.

(3) Switchback occurrence properties.

$P_{SB}$: spatial filling factor of switchbacks in the solar wind on hourly timescales. The data are divided into 1-hour intervals, and $N_i$ denotes the number of switchbacks in the $i$th interval, $I_i$, beginning at $t=t_{b,i}$. The path length of each interval in the switchback frame is $
\Delta s_i=\int_{t_{b,i}}^{t_{b,i+1}} |\boldsymbol{V}_{cross}|dt
$. The switchback occurrence fraction in a set of intervals, $G$, is then defined as
$P_{SB}=\frac{\sum_{N_i>0,I_i\in G}\Delta s_i}{\sum_{I_i\in G} \Delta s_i}\times 100\%$. This quantity represents the global probability of encountering switchback-rich regions.

$\nu_{SB}$: spatial occurrence rate of switchbacks within switchback-containing intervals \citep[e.g.][]{tenerani2021evolution,pecora2022magnetic}, defined as
$\langle\nu_{SB}\rangle=\frac{\sum_{N_i>0, I_i\in G} N_i}{\sum_{N_i>0, I_i\in G} \Delta s_i}$ for $G$ as a set of intervals. It characterizes the local density of switchbacks within active patches.

(4) Switchback size properties.

$\tau_{SB}$: duration of a switchback event.

$D_{SB}$: spatial scale (spacecraft intercept length) of a switchback, defined as
$D_{SB}=\int_{t_{b,SB}}^{t_{b,SB}+\tau_{SB}} |\boldsymbol{V}_{cross}|dt$,
where $t_{b,SB}$ is the onset time of the switchback event (see Figure \ref{fig:8}m).

As shown in Figure \ref{fig:3}, the heliocentric distance at perihelion in E21 ($\sim$10$R_\odot$) is significantly smaller than that in E01 ($\sim$35$R_\odot$), resulting in a lower background solar wind Alfvén Mach number $M_A$, which can even drop below unity. We also identify several switchbacks embedded in sub-Alfvénic solar wind \citep{sioulas2026generation}, including events near perihelion in E21 (Figure \ref{fig:3}b). Among all 4982 switchbacks in E01--E24, 48 are observed under sub-Alfvénic conditions. Consistent with the encounter-to-encounter statistics, both occurrence metrics, $P_{SB}$ and $\nu_{SB}$, are significantly lower in E21 (solar minimum) than in E01 (solar maximum). More generally, switchback properties—including Alfvénicity, occurrence rate, and characteristic size—vary systematically with heliocentric distance. In addition, for different background solar wind conditions characterized by $M_A$, radial velocity $V_R$, and crossing angle $\theta_{VB}$, these properties may exhibit distinct behaviors and anisotropy. To quantify the dependence of switchback properties on the background solar wind, we perform a statistical analysis based on the constructed catalog, with the results presented in the following sections.

\subsection{Radial evolution of switchback properties\label{ssec:rev}}
\begin{figure*}[ht!]
\plotone{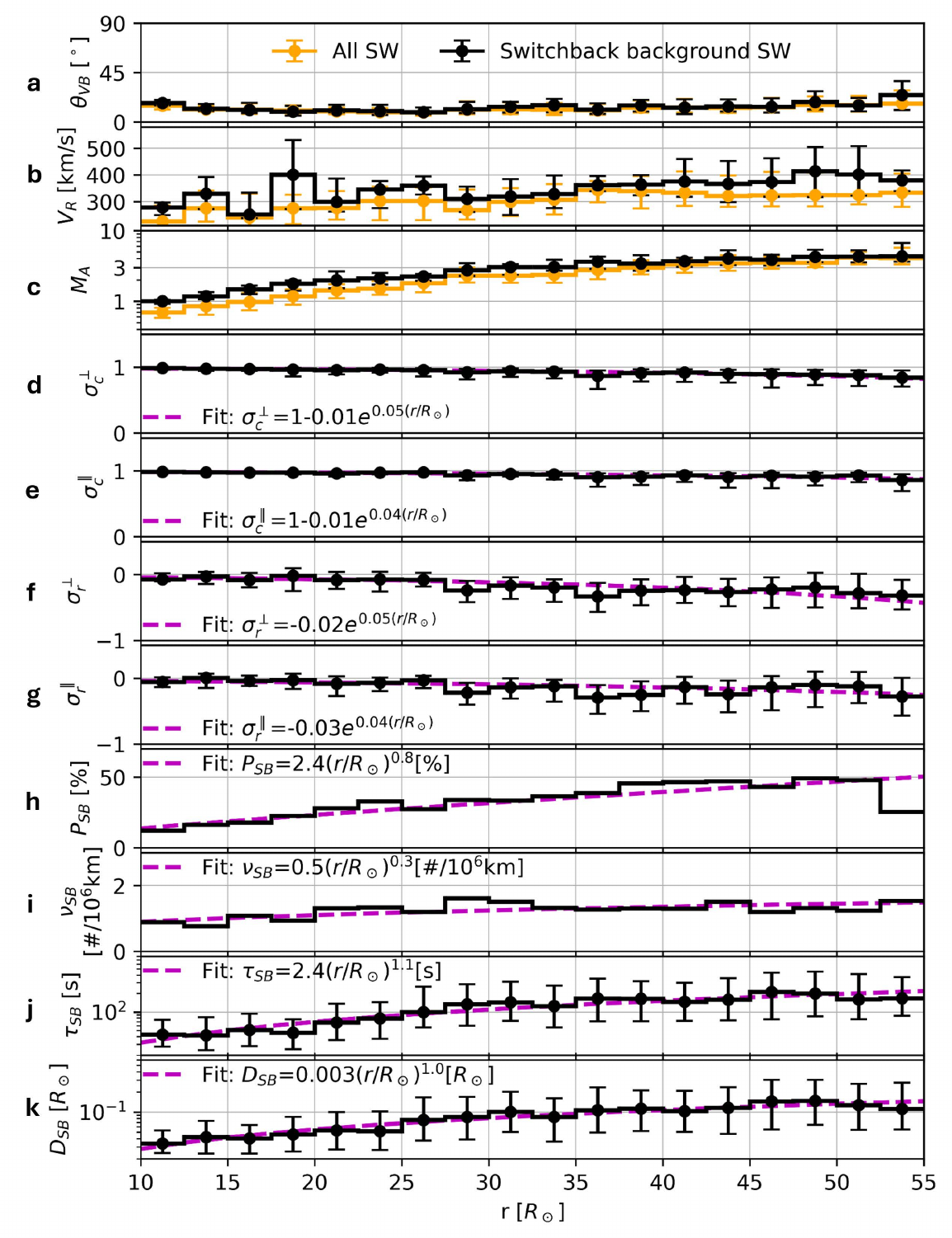}
\caption{Radial evolution of switchback properties and the associated background solar wind over $10<r<55R_\odot$. Panels (a)–(c) show $\theta_{VB}$ (a), $V_R$ (b), and $M_A$ (c) for the background solar wind, comparing intervals containing switchbacks only (black) with all PSP observations in E01–E24 (orange). Panels (d)–(g) present switchback Alfvénicity parameters, including $\sigma_c^\perp$ (d), $\sigma_c^\parallel$ (e), $\sigma_r^\perp$ (f), and $\sigma_r^\parallel$ (g). Panels (h)–(k) show occurrence properties, $P_{SB}$ (h) and $\nu_{SB}$ (i), and size properties, $\tau_{SB}$ (j) and $D_{SB}$ (k). Magenta dashed lines indicate the best-fit radial trends. In panels (a)–(g), (j), and (k), dots represent median values in each radial bin, and error bars indicate the 25th–75th percentile range.
\label{fig:4r}}
\end{figure*}

To investigate the radial evolution of switchback properties, we divide the heliocentric distance range $10<r<55R_\odot$ into 18 equally spaced bins and compute statistical quantities in each bin using PSP observations from E01--E21, as shown in Figure \ref{fig:4r}. To ensure statistical reliability, each radial bin is required to contain more than 30 hr of PSP observations for panels (a)–(c) and (h)–(i), and more than 30 switchbacks for panels (d)–(g) and (j)–(k).

Figures \ref{fig:4r}a–\ref{fig:4r}c show the radial dependence of $\theta_{VB}$, $V_R$, and $M_A$ for intervals containing switchbacks (black), compared with all background solar wind intervals (orange). The angle $\theta_{VB}$ remains nearly constant at $\sim20^\circ$ across the radial range, indicating that PSP predominantly samples outward-propagating Alfvénic fluctuations along trajectories nearly aligned with the background magnetic field, independent of heliocentric distance. The consistently small $\theta_{VB}$ reflects the dominance of the parallel components $V_{PSP}^{\parallel}$ and $V_A$ in $V_{cross}^{\parallel}$, compared with the relatively small perpendicular contribution $V_{PSP}^{\perp}$. In contrast, both $V_R$ and $M_A$ are systematically higher in the background solar wind associated with switchbacks than in the overall solar wind population, suggesting that switchbacks preferentially occur in faster and higher-Alfvén-Mach-number solar wind streams.

Figures \ref{fig:4r}d–\ref{fig:4r}k present the radial evolution of switchback Alfvénicity (d–g), occurrence rate (h–i), and size (j–k), with best-fit trends shown as magenta dashed lines. Both normalized cross helicity $\sigma_c$ and normalized residual energy $\sigma_r$ decrease with increasing heliocentric distance in both parallel and perpendicular components, indicating a gradual reduction of switchback Alfvénicity as the solar wind evolves from strongly Alfvénic fluctuations toward more magnetically dominated turbulence. The similarity between the parallel and perpendicular components suggests that the highly imbalanced turbulence has undergone substantial nonlinear evolution, moving well beyond the simple Alfvén-wave regime.

The occurrence metrics $P_{SB}$ and $\nu_{SB}$ increase with heliocentric distance, following approximate power-law scalings $P_{SB}\propto (r/R_\odot)^{0.8}$ and $\nu_{SB}\propto (r/R_\odot)^{0.3}$. At $r=55R_\odot$, their values are approximately 3.7 and 1.7 times higher than those at $r=10R_\odot$, respectively. This trend suggests that a significant fraction of switchbacks may be generated locally during solar wind expansion between $10<r<55R_\odot$. The characteristic switchback duration and spatial scale also increase with heliocentric distance, following approximate power-law relationships $\tilde{\tau}_{SB}\propto (r/R_\odot)^{1.1}$ and $\tilde{D}_{SB}\propto (r/R_\odot)^{1.0}$, suggesting a quasi linear correlation with the heliocentric distance.


\subsection{Impact of background solar wind \text{Alfv\'en} Mach number \label{ssec:rMA}}

\begin{figure*}[ht!]
\plotone{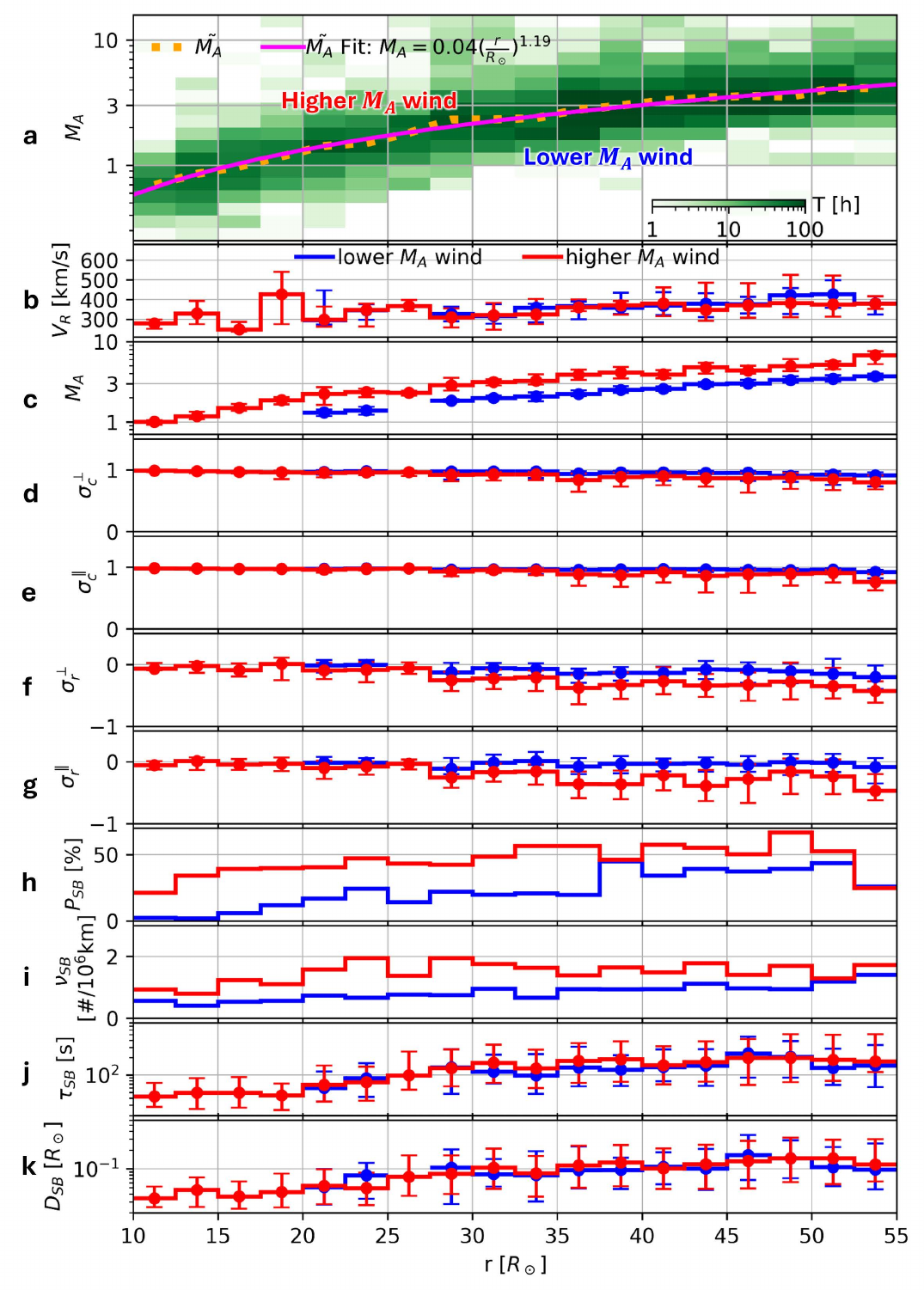}
\caption{Radial evolution of switchback properties and the associated background solar wind, separated by low- and high-$M_A$ regimes over $10<r<55R_\odot$. (a) Joint distribution of PSP observation time (T) with valid level-3 electron density measurements from RFS, shown as a function of heliocentric distance $r$ (horizontal axis) and background solar wind Alfvén Mach number $M_A$ (vertical axis). The median $M_A$ values in each radial bin, $\tilde{M_A}$ (orange dotted line), are fitted with a power-law relation (magenta line), which is used to define the threshold separating low- and high-$M_A$ wind. (b)–(k) Same quantities as in Figure \ref{fig:4r}b–\ref{fig:4r}k, shown separately for low-$M_A$ (blue) and high-$M_A$ (red) solar wind.
\label{fig:4k}}
\end{figure*}

To investigate the influence of the solar wind Alfvén Mach number $M_A$ on the radial evolution of switchback properties, we separate the solar wind into low- and high-$M_A$ regimes using all PSP observations from E01–E24. Figure \ref{fig:4k}a shows the joint distribution of PSP observation time in $r$–$M_A$ space. The median values of $M_A$ in each radial bin, $\tilde{M_A}$, are well described by a power-law relation $\tilde{M_A}(r)=0.04(r/R_\odot)^{1.19}$. Solar wind intervals are then classified as high-$M_A$ ($M_A>\tilde{M_A}(r)$) or low-$M_A$ ($M_A<\tilde{M_A}(r)$) at a given heliocentric distance.

Figures \ref{fig:4k}b–\ref{fig:4k}k present the radial evolution of switchback and background solar wind properties for the high- and low-$M_A$ regimes using the same method as in Figure \ref{fig:4r}b–\ref{fig:4r}k. The radial velocity $V_R$ in the low-$M_A$ wind is generally comparable to, and in some radial ranges slightly higher than, that in the high-$M_A$ wind. By construction, $M_A$ in the high-$M_A$ regime is approximately twice that in the low-$M_A$ regime at the same heliocentric distance. Both $\sigma_c$ and $\sigma_r$ decrease more rapidly with increasing $r$ in the high-$M_A$ wind, indicating a lower level of Alfvénicity compared with the low-$M_A$ wind at the same distance. In terms of occurrence properties, $P_{SB}$ is typically $\sim20\%$ higher in the high-$M_A$ wind, while $\nu_{SB}$ is approximately a factor of two larger than in the low-$M_A$ wind, suggesting that switchback-rich patches preferentially occur in higher-$M_A$ solar wind. In contrast, the characteristic switchback sizes ($\tau_{SB}$ and $D_{SB}$) are comparable between the two regimes at the same heliocentric distance, indicating a weak dependence of switchback size on $M_A$.

\begin{figure*}[ht!]
\plotone{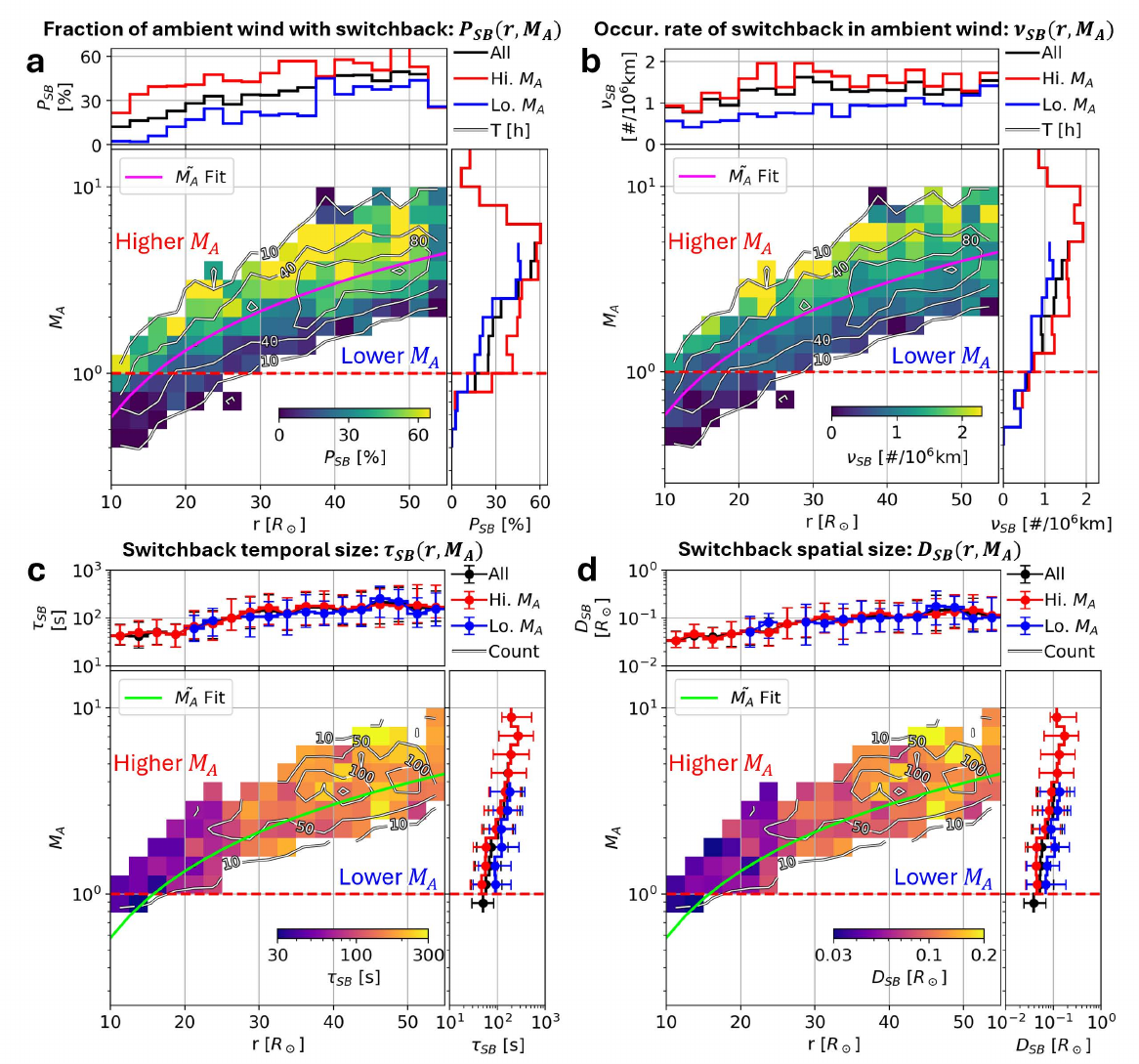}
\caption{Distribution of switchback occurrence rate and size as a function of heliocentric distance and background solar wind Alfvén Mach number. Panels (a) and (b) show the spatial filling factor of switchbacks in 1-hour cadence, $P_{SB}$, and the switchback spatial occurrence rate, $\nu_{SB}$, respectively, plotted in $r$–$M_A$ space, where $r$ is the heliocentric distance and $M_A$ is the Alfvén Mach number. White contours indicate the two-dimensional distribution of PSP observation time with valid level-3 electron density measurements from RFS, $T$. The magenta curve represents the fitted radial dependence of the median $M_A$, $\tilde{M_A}(r)$, which separates low- and high-$M_A$ solar wind regimes. One-dimensional distributions of $P_{SB}$ and $\nu_{SB}$ as functions of $r$ or $M_A$ are shown in the marginal panels for all solar wind (black), high-$M_A$ solar wind (red), and low-$M_A$ solar wind (blue). Panels (c) and (d) show the distributions of the median switchback duration $\tau_{SB}$ and spatial scale $D_{SB}$ using the same format as in (a) and (b). The lime curve denotes $\tilde{M_A}(r)$, and white contours represent the switchback counts. Error bars in the marginal distributions indicate the interquartile range (25th–75th percentile).
\label{fig:4}}
\end{figure*}

\begin{figure*}[ht!]
\plotone{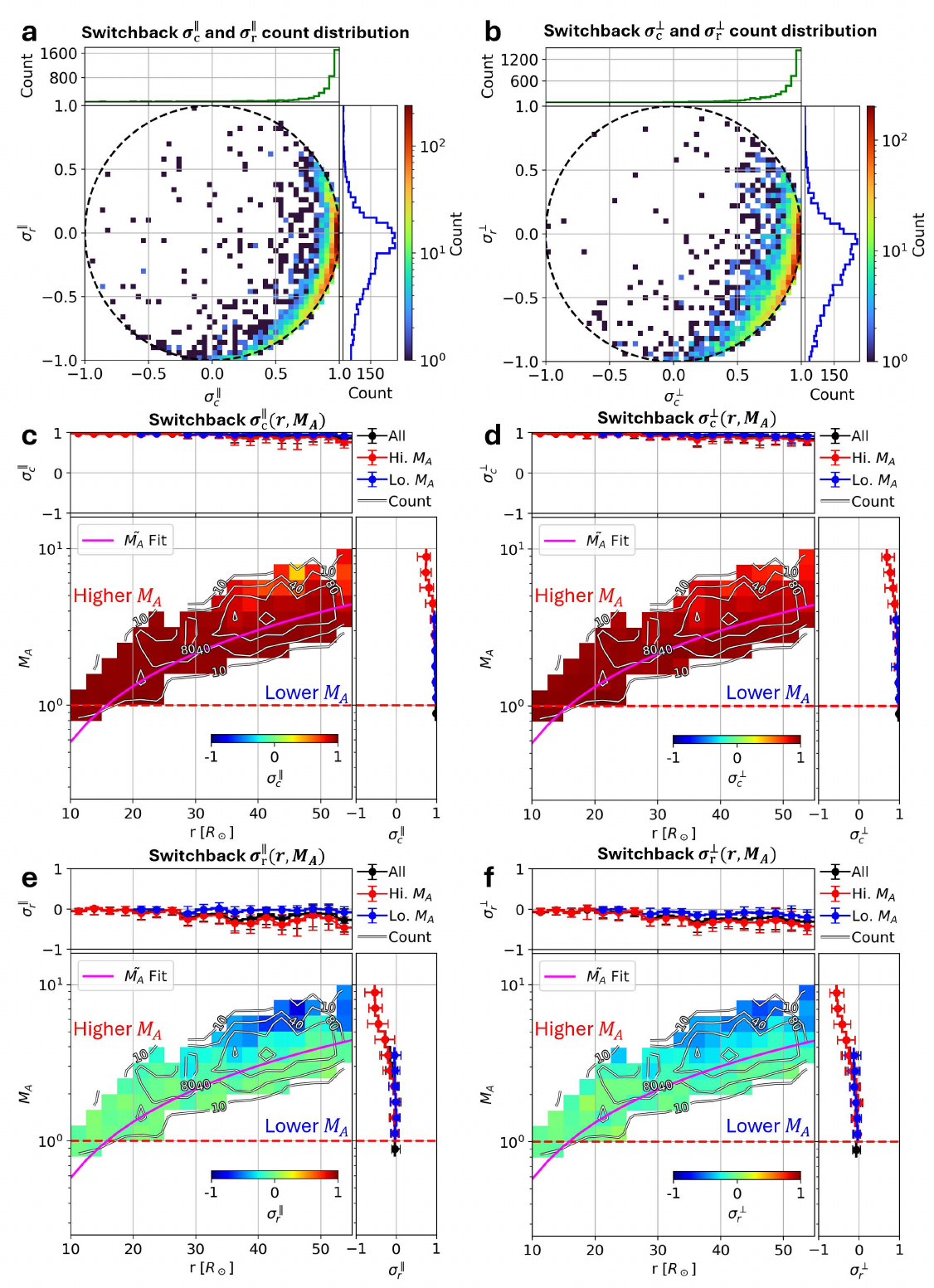}
\caption{Distribution of switchback Alfvénicity properties as a function of heliocentric distance and background solar wind Alfvén Mach number. (a) The switchback count distribution in the $\sigma_c^\parallel$–$\sigma_r^\parallel$ plane. Marginal one-dimensional distributions of counts versus $\sigma_c^\parallel$ and $\sigma_r^\parallel$ are shown in the upper and right panels, respectively. (c) and (e) The distributions of $\sigma_c^\parallel$ and $\sigma_r^\parallel$, respectively, using the same format as in Figures \ref{fig:4}a. White contours indicate the switchback count distribution. Error bars in the marginal distributions represent the interquartile range (25th–75th percentile). (b), (d), and (f) The corresponding results for $\sigma_c^\perp$ and $\sigma_r^\perp$ in the same format as panels (a), (c), and (e).
\label{fig:5}}
\end{figure*}

To better quantify the influence of the Alfvén Mach number $M_A$, we present two-dimensional distributions of switchback properties in $r$–$M_A$ space in Figures \ref{fig:4} and \ref{fig:5}. Figures \ref{fig:4}a and \ref{fig:4}b show the distributions of switchback occurrence properties $P_{SB}$ and $\nu_{SB}$, respectively. To ensure statistical reliability, each two-dimensional bin is required to contain more than 10 hr of PSP observations, while the marginal one-dimensional distributions require at least 30 hr of data. The same $M_A$ threshold derived from the fitted $\tilde{M_A}(r)$ relation in Figure \ref{fig:4k}a is used to separate low- and high-$M_A$ solar wind. In the two-dimensional distributions, all bins with $P_{SB}>50\%$ or $\nu_{SB}>2\times10^{-6}\mathrm{km}^{-1}$ are found exclusively in the high-$M_A$ regime. Figures \ref{fig:4}c and \ref{fig:4}d show the corresponding distributions of the median switchback temporal scale $\tau_{SB}$ and spatial scale $D_{SB}$ in $r$–$M_A$ space. At a given heliocentric distance, both $\tau_{SB}$ and $D_{SB}$ exhibit only weak dependence on $M_A$, indicating that switchback size is only weakly correlated with the background Alfvén Mach number.

Switchback count distributions in the $\sigma_c$–$\sigma_r$ space for the parallel and perpendicular components during E01–E24 are shown in Figures \ref{fig:5}a and \ref{fig:5}b. The nearly identical median values, $\sigma_c^\perp\sim0.92$ with $\sigma_c^\parallel\sim0.94$, and $\sigma_r^\perp\sim-0.2$ with $\sigma_r^\parallel\sim-0.15$, indicate a strongly imbalanced, magnetically dominated, and predominantly sunward-propagating nonlinear Alfvénic state for typical switchbacks over $10<r<55R_\odot$. Figures \ref{fig:5}c–\ref{fig:5}f present the $r$–$M_A$ distributions of the four Alfvénicity parameters, together with comparisons between high-$M_A$ (blue) and low-$M_A$ (red) solar wind, analogous to Figure \ref{fig:4}. Near $r\sim10R_\odot$, switchbacks exhibit high Alfvénicity, with $\sigma_c\sim1$ and $\sigma_r\sim0$. With increasing heliocentric distance, both $\sigma_c$ and $\sigma_r$ decrease, indicating a progressive reduction in Alfvénicity. At $r\gtrsim30R_\odot$, these quantities further decrease with increasing $M_A$, suggesting a more rapid decay of Alfvénicity in high-$M_A$ solar wind. The similar radial trends and comparable behavior between the parallel and perpendicular components suggests high correlation between them. Based on stable magnetic magnitude property inside switchback, $0\simeq\langle\delta (B^2)\rangle=2\langle B_\parallel\rangle\langle\delta B_\parallel\rangle+2\langle B_\perp\rangle\langle\delta B_\perp\rangle$, which leads to $\langle\delta B_\parallel\rangle/\langle\delta B_\perp\rangle=-\langle B_\perp\rangle/\langle B_\parallel\rangle$. Therefore, the high correlation of $\langle\delta B_\parallel\rangle$ and $\langle\delta B_\perp\rangle$ indicates a high correlation of $\langle B_\perp\rangle$ and $\langle B_\parallel\rangle$ inside a switchback. A possible explanation is that the local mean magnetic field direction within a switchback, $\boldsymbol{B}_{0,\mathrm{in}}$, may differ from the background field direction in the ambient solar wind, $\boldsymbol{B}_{0,\mathrm{out}}$. As a result, fluctuations that are primarily perpendicular to $\boldsymbol{B}_{0,\mathrm{in}}$ can project onto both parallel and perpendicular components defined with respect to $\boldsymbol{B}_{0,\mathrm{out}}$ in this analysis.

\subsection{Impact of background solar wind speed \label{ssec:rVR}}

\begin{figure*}[ht!]
\plotone{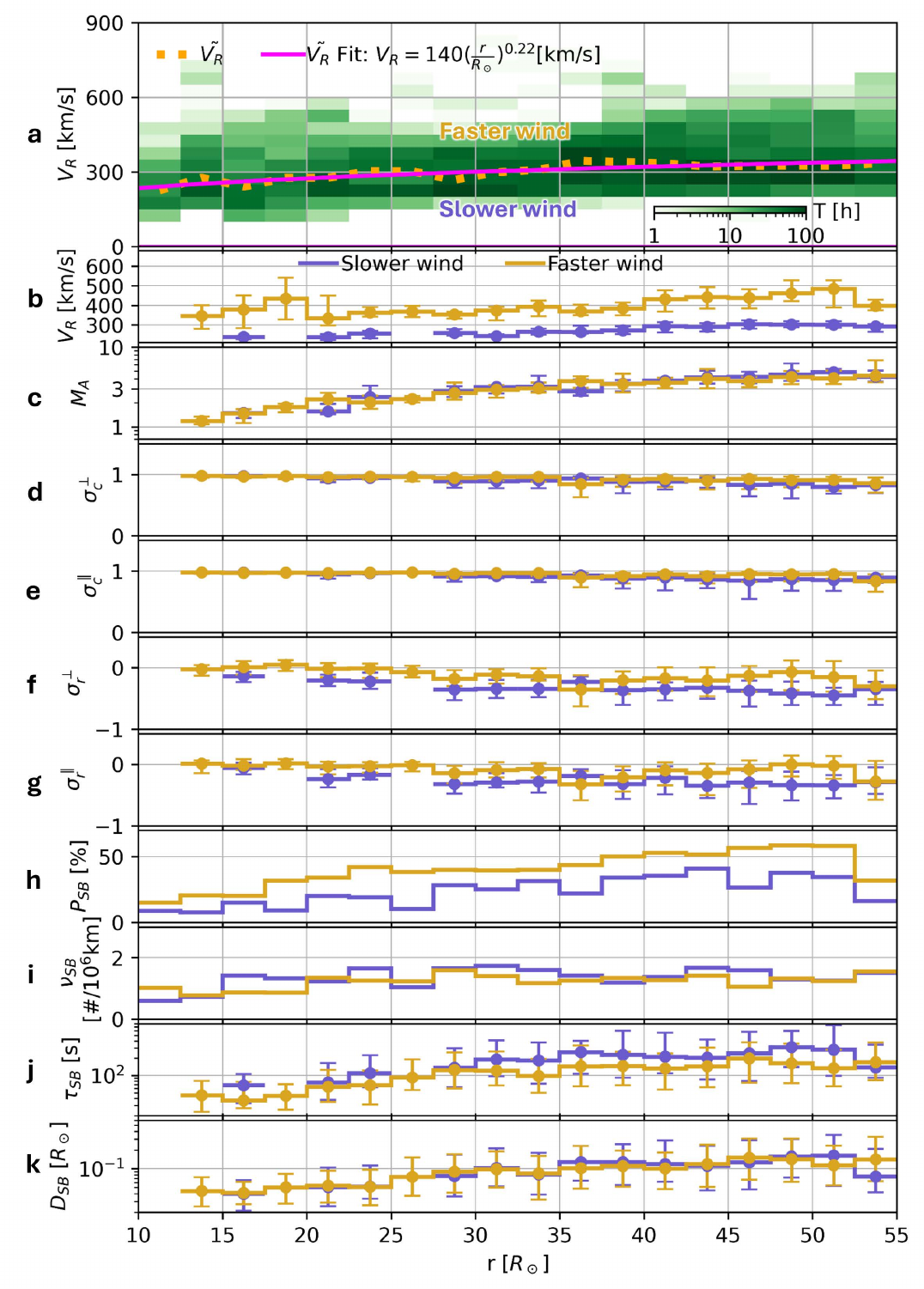}
\caption{Radial evolution of switchback properties and the associated background solar wind, separated into slow- and fast-wind regimes over $10<r<55R_\odot$. (a) Joint distribution of PSP observation time (T) with valid level-3 electron density measurements from RFS, shown as a function of heliocentric distance $r$ (horizontal axis) and background solar wind radial velocity $V_R$ (vertical axis). The median $V_R$ values in each radial bin, $\tilde{V_R}$ (orange dotted line), are fitted with a power-law relation (magenta line), which is used to define the threshold separating slow and fast solar wind. (b)–(k) Same quantities as in Figure \ref{fig:4r}b–\ref{fig:4r}k, shown separately for slow-wind (slate-blue) and fast-wind (goldenrod) regimes.
\label{fig:6}}
\end{figure*}

To investigate the influence of background solar wind speed $V_R$ on switchback properties, we use a power-law fit to the median solar wind speed in each radial bin, $\tilde{V_R}$, as shown in Figure \ref{fig:6}a. The fitted relation is $\tilde{V_R}=140(r/R_\odot)^{0.22}$ km s$^{-1}$, with $\tilde{V_R}(1,\mathrm{AU})\approx460$ km s$^{-1}$, which is consistent with the typical fast/slow wind separation at 1 AU. Solar wind intervals are then classified into fast- and slow-wind regimes relative to this threshold. The radial velocity in fast wind is typically more than 150 km s$^{-1}$ higher than that in slow wind at the same heliocentric distance, as shown in Figure \ref{fig:6}b, while the corresponding $M_A$ values remain comparable between the two regimes (Figure \ref{fig:6}c).

Figures \ref{fig:6}d–\ref{fig:6}g show the radial evolution of switchback Alfvénicity properties. In the fast wind, both $\sigma_c$ and $\sigma_r$ remain close to 1 and 0, respectively, whereas they decrease more rapidly with increasing heliocentric distance in the slow wind. This indicates that fast solar wind maintains a higher level of Alfvénicity, consistent with near-Sun turbulence observations \citep{sioulas2025propagation} and theoretical expectations from turbulence-driven models \citep{meyrand2025reflection}.

Figures \ref{fig:6}h–\ref{fig:6}k compare the occurrence rate and size properties between fast and slow wind as a function of $r$. The occurrence fraction $P_{SB}$ is approximately twice as large in fast wind, indicating a higher probability of encountering switchback-rich patches. In contrast, the local occurrence rate $\nu_{SB}$ is similar between fast and slow wind, suggesting that background velocity primarily modulates the large-scale patchiness rather than the local switchback density within patches. Regarding size, the temporal scale $\tau_{SB}$ is larger in slow wind, while the spatial scale $D_{SB}$ is nearly identical between the two regimes. This implies a weak dependence of switchback spatial extent on $V_R$, with shorter durations in fast wind arising primarily from higher advection speed at comparable spatial scales.

\subsection{Impact of solar activity \label{ssec:rSA}}

\begin{figure*}[ht!]
\plotone{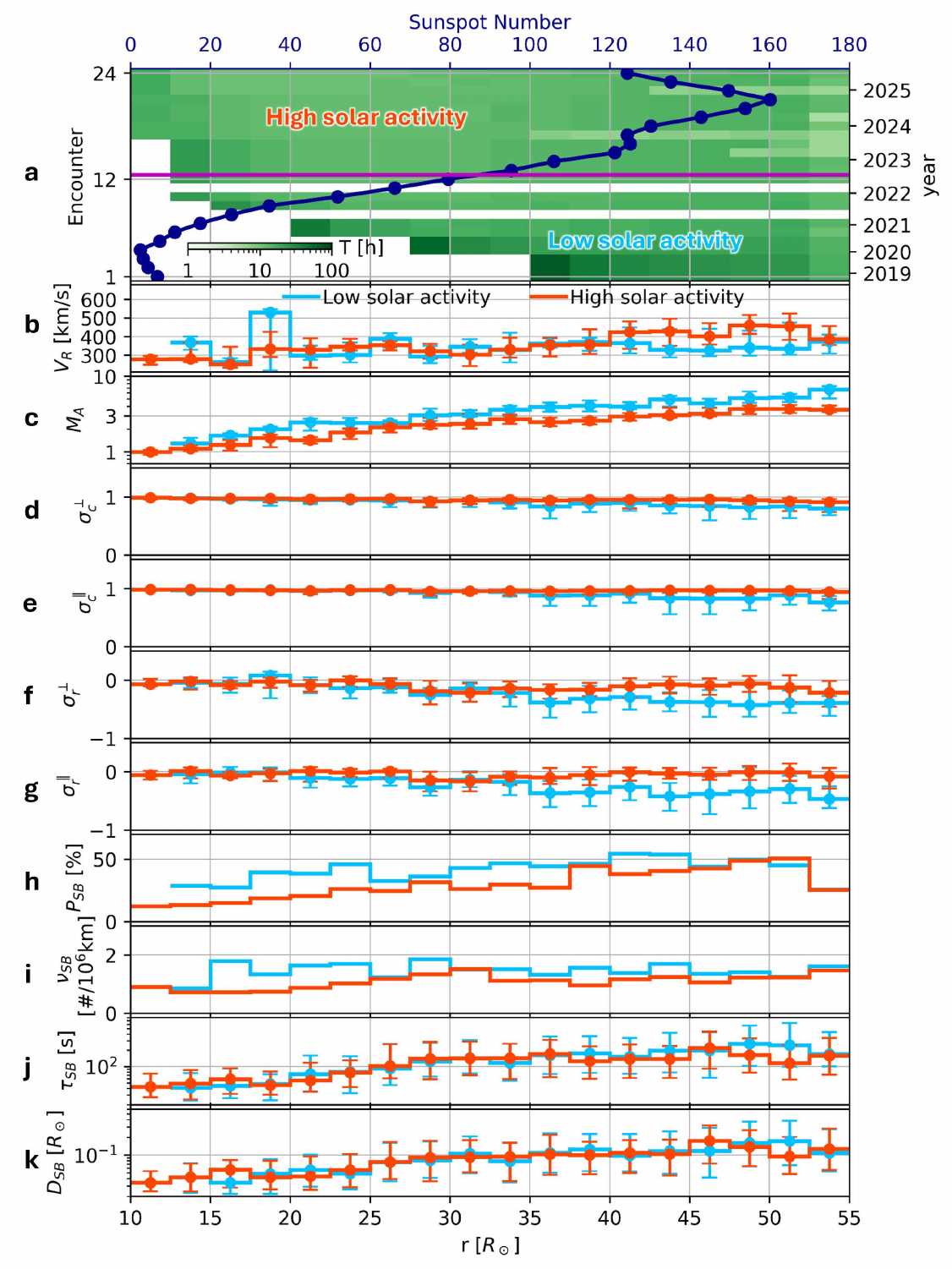}
\caption{Radial evolution of switchback properties and the associated background solar wind under low- and high-solar-activity conditions over $10<r<55R_\odot$. (a) Joint distribution of PSP observation time (T) with valid level-3 electron density measurements from RFS, shown as a function of heliocentric distance $r$ (lower horizontal axis) and PSP encounter number (left vertical axis). The corresponding sunspot number (upper horizontal axis), mapped to time (right vertical axis), is shown as connected dark blue points. Encounters E01–E12 correspond to low solar activity (sunspot number $<90$), while E13–E24 correspond to high solar activity (sunspot number $>90$), separated by a magenta line at encounter number 12.5. (b)–(k) Same quantities as in Figure \ref{fig:4r}b–\ref{fig:4r}k, shown separately for low-activity (deep-sky-blue) and high-activity (orange-red) intervals.
\label{fig:7}}
\end{figure*}

Solar activity varies significantly from E01 in 2018 (near solar minimum) to E24 in 2025 (near solar maximum). As an indicator of solar activity, the monthly mean sunspot number is shown for each encounter as the dark blue curve in Figure \ref{fig:7}a. Using a threshold of 90, the data are divided into a low-activity group (E01–E12, 2018–2022; sunspot number $<90$) and a high-activity group (E13–E24, 2022–2025; sunspot number $>90$).

Figures \ref{fig:7}d–\ref{fig:7}k present the switchback Alfvénicity, occurrence rate, and size properties during low (deep sky blue) and high (orange-red) solar activity. Both $\sigma_c$ and $\sigma_r$ are higher during high solar activity, with values closer to 1 and 0, respectively, indicating a higher degree of Alfvénicity during solar maximum compared to solar minimum. In contrast, both $P_{SB}$ and $\nu_{SB}$ are larger during low solar activity, particularly for $r<40R_\odot$. Around $r\sim20R_\odot$, $P_{SB}$ and $\nu_{SB}$ during solar minimum are nearly twice those during solar maximum, while this difference decreases to nearly unity at $r\sim50R_\odot$. The characteristic switchback size, $\tau_{SB}$ and $D_{SB}$, shows no significant dependence on solar activity.

To investigate the origin of these differences, we examine the radial profiles of the background solar wind speed $V_R$ and Alfvén Mach number $M_A$ in Figures \ref{fig:7}b and \ref{fig:7}c. During high solar activity, $M_A$ is systematically lower, consistent with the higher Alfvénicity and lower occurrence rates observed in low-$M_A$ solar wind, as discussed in Section \ref{ssec:rMA}. The radial velocity $V_R$ is similar for $r<35R_\odot$, but becomes relatively lower during low solar activity at larger distances, suggesting a higher fraction of the ecliptic faster wind during solar maximum. In contrast, faster wind during high solar activity at $r>40R_\odot$ enhances $P_{SB}$, partially compensating for the lower occurrence expected in low-$M_A$ conditions during high solar activity. Other properties show no comparable dependence on $V_R$, consistent with Section \ref{ssec:rVR}. Overall, these results suggest that the influence of solar activity on switchbacks is primarily mediated through its modulation of the background $M_A$ distribution and the resulting occurrence of switchback-rich solar wind intervals.

\subsection{The anisotropy of switchback occurrence rate and size \label{ssec:ralpha}}

\begin{figure*}[ht!]
\plotone{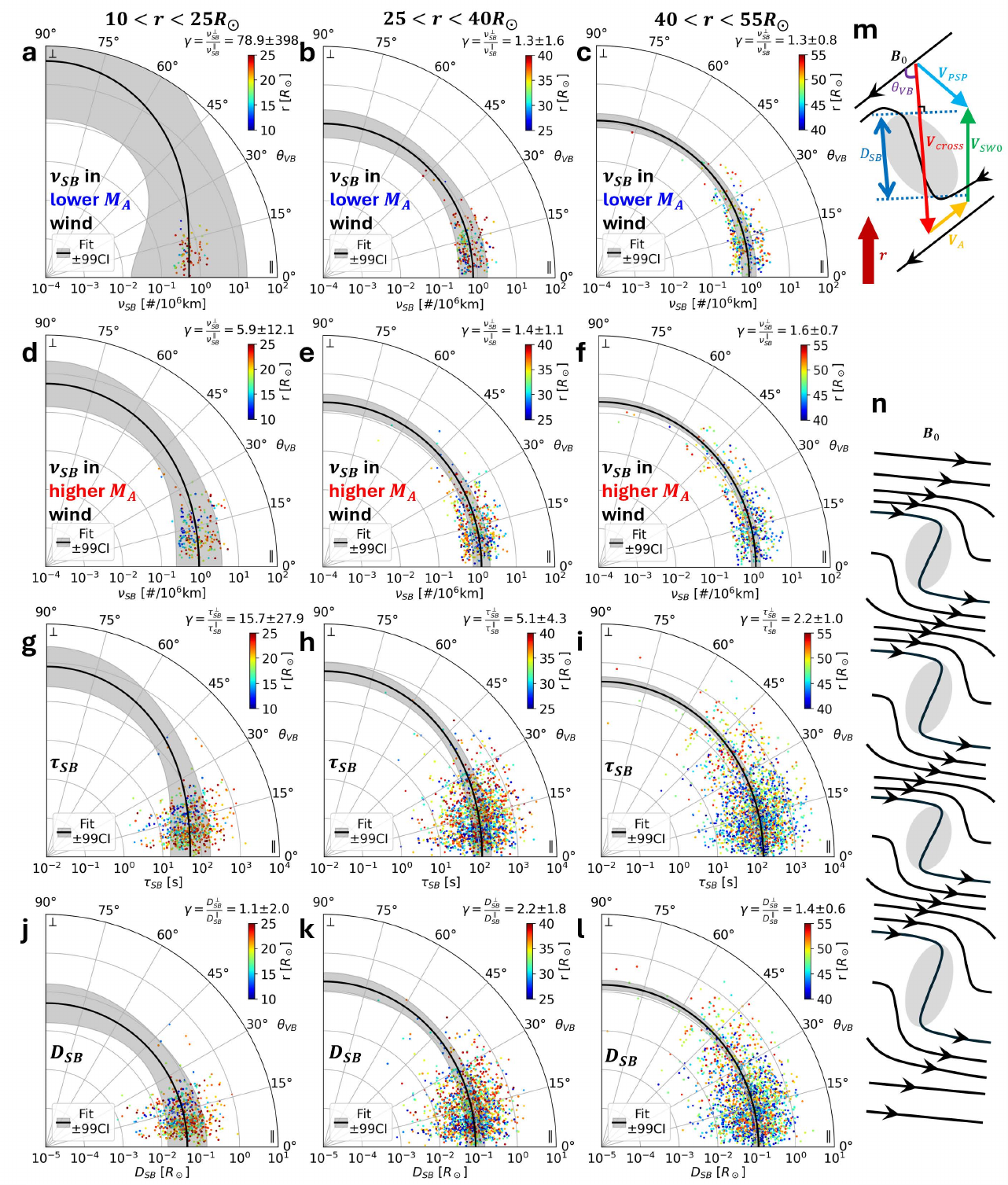}
\caption{Switchback occurrence rate and size anisotropy relative to the background magnetic field direction. (a)–(l) Polar distributions of switchbacks as a function of $\theta_{VB}$ and either hourly occurrence rate $\nu_{SB}$, temporal size $\tau_{SB}$, or spatial size $D_{SB}$ for low-$M_A$ wind (a–c), high-$M_A$ wind (d–f), and three heliocentric distance ranges: (10–25)$R_\odot$ (a,d,g,j), (25–40)$R_\odot$ (b,e,h,k), and (40–55)$R_\odot$ (c,f,i,l). Black curves show the fitted angular dependence $X(\theta_{VB})=10^{a_0-a_1\cos(2\theta_{VB})}$ with 99$\%$ confidence intervals. The anisotropy ratio $\gamma=X^\perp/X^\parallel$ is shown in each panel, where $X^\perp=X(90^\circ)$ and $X^\parallel=X(0^\circ)$.
(m) schematic illustrations of $\theta_{VB}$, defined as the unsigned angle between the relative PSP velocity $\boldsymbol{V}_{\mathrm{cross}}$ (red arrow) and the background magnetic field $\boldsymbol{B}_0$ (black arrow) in the switchback reference frame. (n) A conceptual model of switchback magnetic field topology in a patch, respectively. Multiple switchbacks with aspect ratio $\gamma\sim1.5$ are aligned preferentially in the perpendicular direction within a patch. Switchbacks with magnetic field polarity reversal relative to $\boldsymbol{B}_0$ are highlighted in gray.
\label{fig:8}}
\end{figure*}

Using the switchback catalog, we analyze the anisotropy of switchback occurrence rate and size with respect to the background magnetic field direction in the switchback frame, parameterized by $\theta_{VB}$ as defined in Section \ref{sec:result}. Figure \ref{fig:8} shows switchbacks in polar coordinates, where the radial axis represents the occurrence rate $\nu_{SB}$ in every 1-hour intervals with switchbacks (Figures \ref{fig:8}a–\ref{fig:8}f), temporal size $\tau_{SB}$ (Figures \ref{fig:8}g–\ref{fig:8}i), or spatial size $D_{SB}$ for all the measurable switchbacks in our catalog (Figures \ref{fig:8}j–\ref{fig:8}l), and the angular coordinate corresponds to $\theta_{VB}$. A logarithmic scale is used for the radial axis due to the wide dynamic range of these quantities.

To examine radial effects, results are divided into three heliocentric distance ranges: (10–25)$R_\odot$, (25–40)$R_\odot$, and (40–55)$R_\odot$, with each panel color-coded by $r$. The influence of radial evolution within each bin is considered secondary compared with angular dependence. Since $\nu_{SB}$ is systematically higher in high-$M_A$ wind at the same $r$, its anisotropy is shown separately for low- and high-$M_A$ regimes in Figures \ref{fig:8}a–\ref{fig:8}c and \ref{fig:8}d–\ref{fig:8}f, respectively.

The anisotropy of each quantity $X$ ($\nu_{SB}$, $\tau_{SB}$, or $D_{SB}$) is quantified using the fitting form
\begin{equation}
X(\theta_{VB}) = 10^{a_0 - a_1 \cos(2\theta_{VB})}.
\end{equation}
The fitted curves are shown as thick black lines in Figures \ref{fig:8}a–\ref{fig:8}l, with shaded regions indicating $99\%$ confidence intervals. The fitting uncertainty decreases with increasing $r$, primarily due to improved sampling of quasi-perpendicular events at larger distances. Closer to the Sun, the larger magnitude of the Alfvén velocity relative to $\boldsymbol{V}_{PSP}$ and $\boldsymbol{V}_{SW}$ causes $\boldsymbol{V}_{cross}$ to be more strongly aligned with the radial direction, resulting in fewer high-$\theta_{VB}$ occurrences.

\begin{deluxetable*}{ccccccc}
\digitalasset
\tablewidth{0pt}
\tablecaption{$\nu_{SB}$, $\tau_{SB}$, and $D_{SB}$ anisotropy based on fitting results \label{tab:2}}
\tablehead{
\colhead{$X$} & \colhead{$r$ range [$R_\odot$]} & \colhead{$a_0$} & \colhead{$a_1$} & \colhead{$X^\parallel=10^{a_0-a_1}$} & \colhead{$X^\perp=10^{a_0+a_1}$} & \colhead{$\gamma=X^\perp/X^\parallel$}  
}
\startdata
$\nu_{SB}$ in    & $10\sim25$ & $0.66^{\pm1.02}$  & $0.95^{\pm1.10}$ & $0.5_{-0.5}^{+15.7}$ & $41_{-39}^{+1238}$  & $79^{\pm398}$  \\
lower $M_A$ wind & $25\sim40$ & $-0.06^{\pm0.24}$ & $0.06^{\pm0.27}$ & $0.8_{-0.4}^{+1.0}$  & $1.0_{-0.6}^{+1.3}$ & $1.3^{\pm1.6}$ \\
$[\#/10^6km]$    & $40\sim55$ & $0.01^{\pm0.12}$  & $0.05^{\pm0.14}$ & $0.9_{-0.3}^{+0.5}$  & $1.1_{-0.4}^{+0.6}$ & $1.3^{\pm0.8}$ \\
\tableline
$\nu_{SB}$ in    & $10\sim25$ & $0.36^{\pm0.39}$  & $0.38^{\pm0.45}$ & $0.9_{-0.7}^{+2.7}$  & $6_{-4}^{+16}$      & $6^{\pm12}$  \\
higher $M_A$ wind& $25\sim40$ & $0.19^{\pm0.14}$  & $0.08^{\pm0.17}$ & $1.3_{-0.5}^{+0.8}$  & $1.8_{-0.7}^{+1.2}$ & $1.4^{\pm1.1}$ \\
$[\#/10^6km]$    & $40\sim55$ & $0.18^{\pm0.08}$  & $0.11^{\pm0.10}$ & $1.2_{-0.3}^{+0.4}$  & $1.9_{-0.5}^{+0.6}$ & $1.6^{\pm0.7}$ \\
\tableline
                 & $10\sim25$ & $2.30^{\pm0.34}$  & $0.60^{\pm0.39}$ & $51_{-35}^{+116}$    & $799_{-556}^{+1827}$& $16^{\pm28}$   \\
$\tau_{SB}$ [s]  & $25\sim40$ & $2.43^{\pm0.16}$  & $0.35^{\pm0.18}$ & $119_{-51}^{+88}$    & $606_{-258}^{+449}$ & $5.1^{\pm4.3}$ \\
                 & $40\sim55$ & $2.35^{\pm0.08}$  & $0.17^{\pm0.10}$ & $150_{-39}^{+53}$    & $328_{-85}^{+116}$  & $2.2^{\pm1.0}$ \\
\tableline
                 & $10\sim25$ & $-1.32^{\pm0.34}$ & $0.03^{\pm0.38}$ &$0.05_{-0.03}^{+0.10}$&$0.05_{-0.04}^{+0.11}$& $1.1^{\pm2.0}$  \\
$D_{SB}$ [$R_\odot$]&$25\sim40$&$-0.90^{\pm0.16}$ & $0.17^{\pm0.18}$ &$0.08_{-0.04}^{+0.06}$&$0.19_{-0.08}^{+0.14}$& $2.2^{\pm1.8}$ \\
                 & $40\sim55$ & $-0.89^{\pm0.08}$ & $0.07^{\pm0.10}$ &$0.11_{-0.03}^{+0.04}$&$0.15_{-0.04}^{+0.05}$& $1.4^{\pm0.6}$ \\
\enddata
\end{deluxetable*}

For each property and each $r$ range, the fitting result in perpendicular direction as $X^\perp=X(\theta_{VB}=90^\circ)=10^{a_0+a_1}$ is significantly higher than that along parallel direction as $X^\parallel=X(\theta_{VB}=0^\circ)=10^{a_0-a_1}$. We calculate the perpendicular to parallel component aspect ratio as $\gamma=X^\perp/X^\parallel$ based on the fitting result by formula $\gamma=10^{2a_1}$, as shown at upper right of Figure \ref{fig:8}a-\ref{fig:8}i and table \ref{tab:2} with uncertainty in 99$\%$ confidence interval. With smaller uncertainty relative to $\gamma$ value at $r>25R_\odot$, the anisotropy shown at larger $r$ is much reliable. We find the occurrence rate anisotropy keeps around 1.5 no matter in higher or lower $M_A$ wind at $25<r<55R_\odot$. The size anisotropy decreases a little at heliocentric distance from $25<r<40R_\odot$ to $40<r<55R_\odot$ with $\gamma_{\tau_{SB}}$ from 5.1 to 2.2, and $\gamma_{D_{SB}}$ from 2.2 to 1.4. In summary, at $25<r<55R_\odot$, the switchback spatial size or occurrence rate in perpendicular direction is around 1.5 times that in parallel direction relative to background solar wind magnetic direction.

\section{Discussion and Conclusions} \label{sec:D&C}
A comprehensive switchback catalog at heliocentric distances $r<55R_\odot$ is constructed using observations from the first 24 PSP encounters. Potential switchbacks are identified as intervals with local magnetic field deflections exceeding $90^\circ$, and refined using additional constraints on magnetic field magnitude stability and strahl-electron pitch-angle consistency. In this catalog, switchback durations $\tau_{SB}$ span from $\sim1$ s to $\sim10^4$ s, and their occurrence distribution approximately follows a power-law scaling of $\tau_{SB}^{-5/2}$.

Based on statistical analyses of the radial evolution of Alfvénicity, occurrence rate, and size properties, the main results are summarized as follows.

(1) Near-Sun switchback Alfvénicity decreases with heliocentric distance, and the decay rate is modulated by the background solar wind speed and Alfvén Mach number. From $\sim10R_\odot$ near the Alfvén critical point to $55R_\odot$, the normalized cross helicity and residual energy decrease from approximately 1 and 0 to about 0.8 and $-0.3$, respectively, indicating a nonlinear evolution from nearly sunward-propagating Alfvénic fluctuations toward a more imbalanced and magnetically dominated state. The decay rate is faster in slow and high-$M_A$ solar wind, consistent with trends reported for general near-Sun turbulence \citep{sioulas2026generation}. This suggests a similar irreversible Alfvénic turbulence evolution process in the expanding super-Alfvénic solar wind beyond the Alfvén critical point, where the effective evolution timescale depends on both the distance from the critical point and the solar wind speed \citep{shi2021alfvenic}. This mechanism warrants further investigation using combined observations and simulations.

(2) The occurrence rate of near-Sun switchbacks increases with heliocentric distance and varies with the background solar wind speed and Alfvén Mach number. From $\sim10R_\odot$ to $55R_\odot$, the fraction of solar wind containing switchbacks increases from $\sim15\%$ to $\sim50\%$, while the local occurrence rate rises from $\sim0.8$ to $\sim1.3$ per $10^6$ km. These results suggest ongoing generation or accumulation of switchbacks as the solar wind evolves from sub-Alfvénic to super-Alfvénic regions. The occurrence fraction is significantly higher in fast and high-$M_A$ wind, indicating that such conditions are more favorable for switchback generation. The local occurrence rate is also enhanced in high-$M_A$ wind but shows little dependence on solar wind speed, implying that switchback-rich patches are primarily associated with high-$M_A$ source regions. Since PSP trajectories are predominantly quasi-parallel to the background magnetic field when crossing switchbacks, these statistics mainly reflect occurrence variations along the field direction and are interpreted as temporal modulation along the spacecraft path. Our results on the radial evolution of the occurrence rate support multiple switchback generation mechanisms, suggesting that while some switchbacks may originate in the lower corona, a substantial fraction are likely generated locally in the expanding solar wind.  

(3) Solar activity modulates near-Sun switchback Alfvénicity and occurrence rate primarily through its influence on the background solar wind Alfvén Mach number distribution. We find that the median $M_A$ of the switchback-associated background wind during low solar activity is nearly twice that during high solar activity, while the solar wind speed is broadly similar between the two periods, except beyond $r>40R_\odot$, where the wind is faster during high solar activity. At $r<40R_\odot$, the higher $M_A$ in low solar activity leads to reduced Alfvénicity and lower occurrence rates. At $40<r<55R_\odot$, the slower wind during low solar activity further reduces Alfvénicity and partially counteracts the higher-$M_A$ effect on occurrence rate. Overall, the solar cycle dependence is mainly driven by changes in the distribution of source solar wind conditions, particularly the Alfvén Mach number.

(4) The spatial size of near-Sun switchbacks increases approximately linearly with the heliocentric distance, $r$, with median spatial scale of about 0.003$r$. Given that the median PSP crossing angle, $\theta_{VB}$, is $\sim20^\circ$ near the Sun, this scale primarily represents the component parallel to the background magnetic field, which is nearly radial. The spatial size shows no significant dependence on background solar wind speed or Alfvén Mach number, indicating that expansion is not primarily controlled by local source conditions.

(5) Near-Sun switchbacks exhibit clear anisotropy in both occurrence rate and size relative to the background magnetic field direction, with perpendicular values roughly 1.5 times those in the parallel direction. The anisotropy is quantified using the ratio of perpendicular to parallel components derived from fitting results. The occurrence-rate anisotropy is approximately 1.5 for $25<r<55R_\odot$, largely independent of $M_A$, while the spatial-size anisotropy is about 1.4 at $40<r<55R_\odot$. These results indicate an intrinsically anisotropic magnetic-field topology within switchbacks and their associated patches.

Based on the results of this study, several directions for future work can be pursued. First, the solar-disk source regions of the switchback-associated background wind may provide key constraints on the formation and modulation of switchback patches, particularly for those initially generated in sub-Alfvénic conditions \citep{2021ApJ...923..174B,2022ApJ...934..152S}. A statistical investigation of the spatial and temporal mapping between near-Sun switchbacks and their solar-source regions would help clarify this connection.

Second, the radial evolution of near-Sun switchbacks raises further questions regarding the nonlinear evolution of Alfvénic turbulence, especially the physical origin of the observed decay in Alfvénicity and the concurrent increase in occurrence rate. For the ambient solar wind associated with switchbacks, we find enhanced occurrence rates in faster and higher Alfvén Mach number streams. A more detailed comparison of the ambient plasma conditions—such as magnetic field strength, density, and temperature—across different occurrence-rate regimes may help identify the key drivers of switchback generation.

Third, the present results provide evidence for anisotropic switchback topology, including directional dependence of occurrence rate and size, as well as possible differences between the magnetic-field orientations inside and outside individual switchbacks. In future work, these constraints could be used to construct an empirical model of the magnetic-field and plasma structure of near-Sun switchbacks based on our switchback catalog.

Finally, the identification method developed in this study can be extended to other spacecraft datasets, including Solar Orbiter \citep{horbury2020solar} in the inner heliosphere, Wind \citep{lepping1995wind} and ACE \citep{smith1998ace} at 1 AU, and Ulysses \citep{balogh1992magnetic} in the outer heliosphere. Such a multi-mission analysis would enable a more complete characterization of the radial evolution of switchback properties from the Alfvén critical region to several AU.

\begin{acknowledgments}
This work is supported by NASA ECIP \#80NSSC26K0321 and NASA FINESST \#80NSSC26K0478. The instruments of PSP were designed and developed under NASA contract NNN06AA01C. We thank the PSP SWEAP team lead by J.K. and FIELDS team lead by S.B. for use of data. The level-2 magnetometer data can be found at \url{DOI: 10.48322/0yy0-ba92}, the level-3 RFS electron data can be found at \url{DOI: 10.48322/jdj2-1912}, and the level-3 SPAN proton and electron data can be found at \url{DOI: 10.48322/ypyh-s325} and \url{DOI: 10.48322/8ync-7p95}. The switchback catalog, along with their begin and end time,
are uploaded to Zenodo at \url{DOI:10.5281/zenodo.21314515}.
\end{acknowledgments}

\bibliography{sample701}{}

@ARTICLE{2016SSRv..204...49B,
       author = {{Bale}, S.~D. and {Goetz}, K. and {Harvey}, P.~R. and {Turin}, P. and {Bonnell}, J.~W. and {Dudok de Wit}, T. and {Ergun}, R.~E. and {MacDowall}, R.~J. and {Pulupa}, M. and {Andre}, M. and {Bolton}, M. and {Bougeret}, J.-L. and {Bowen}, T.~A. and {Burgess}, D. and {Cattell}, C.~A. and {Chandran}, B.~D.~G. and {Chaston}, C.~C. and {Chen}, C.~H.~K. and {Choi}, M.~K. and {Connerney}, J.~E. and {Cranmer}, S. and {Diaz-Aguado}, M. and {Donakowski}, W. and {Drake}, J.~F. and {Farrell}, W.~M. and {Fergeau}, P. and {Fermin}, J. and {Fischer}, J. and {Fox}, N. and {Glaser}, D. and {Goldstein}, M. and {Gordon}, D. and {Hanson}, E. and {Harris}, S.~E. and {Hayes}, L.~M. and {Hinze}, J.~J. and {Hollweg}, J.~V. and {Horbury}, T.~S. and {Howard}, R.~A. and {Hoxie}, V. and {Jannet}, G. and {Karlsson}, M. and {Kasper}, J.~C. and {Kellogg}, P.~J. and {Kien}, M. and {Klimchuk}, J.~A. and {Krasnoselskikh}, V.~V. and {Krucker}, S. and {Lynch}, J.~J. and {Maksimovic}, M. and {Malaspina}, D.~M. and {Marker}, S. and {Martin}, P. and {Martinez-Oliveros}, J. and {McCauley}, J. and {McComas}, D.~J. and {McDonald}, T. and {Meyer-Vernet}, N. and {Moncuquet}, M. and {Monson}, S.~J. and {Mozer}, F.~S. and {Murphy}, S.~D. and {Odom}, J. and {Oliverson}, R. and {Olson}, J. and {Parker}, E.~N. and {Pankow}, D. and {Phan}, T. and {Quataert}, E. and {Quinn}, T. and {Ruplin}, S.~W. and {Salem}, C. and {Seitz}, D. and {Sheppard}, D.~A. and {Siy}, A. and {Stevens}, K. and {Summers}, D. and {Szabo}, A. and {Timofeeva}, M. and {Vaivads}, A. and {Velli}, M. and {Yehle}, A. and {Werthimer}, D. and {Wygant}, J.~R.},
        title = "{The FIELDS Instrument Suite for Solar Probe Plus. Measuring the Coronal Plasma and Magnetic Field, Plasma Waves and Turbulence, and Radio Signatures of Solar Transients}",
      journal = {\ssr},
         year = 2016,
        month = dec,
       volume = {204},
       number = {1-4},
        pages = {49-82},
          doi = {10.1007/s11214-016-0244-5},
       adsurl = {https://ui.adsabs.harvard.edu/abs/2016SSRv..204...49B}
}

@ARTICLE{2017JGRA..122.2836P,
       author = {{Pulupa}, M. and {Bale}, S.~D. and {Bonnell}, J.~W. and {Bowen}, T.~A. and {Carruth}, N. and {Goetz}, K. and {Gordon}, D. and {Harvey}, P.~R. and {Maksimovic}, M. and {Mart{\'\i}nez-Oliveros}, J.~C. and {Moncuquet}, M. and {Saint-Hilaire}, P. and {Seitz}, D. and {Sundkvist}, D.},
        title = "{The Solar Probe Plus Radio Frequency Spectrometer: Measurement requirements, analog design, and digital signal processing}",
      journal = {Journal of Geophysical Research (Space Physics)},
         year = 2017,
        month = mar,
       volume = {122},
       number = {3},
        pages = {2836-2854},
          doi = {10.1002/2016JA023345},
       adsurl = {https://ui.adsabs.harvard.edu/abs/2017JGRA..122.2836P}
}

@article{Kruparova_2023,
doi = {10.3847/1538-4357/acf572},
url = {https://doi.org/10.3847/1538-4357/acf572},
year = {2023},
month = {oct},
publisher = {The American Astronomical Society},
volume = {957},
number = {1},
pages = {13},
author = {Kruparova, Oksana and Krupar, Vratislav and Szabo, Adam and Pulupa, Marc and Bale, Stuart D.},
title = {Quasi-thermal Noise Spectroscopy Analysis of Parker Solar Probe Data: Improved Electron Density Model for Solar Wind},
journal = {The Astrophysical Journal}
}

@ARTICLE{2016SSRv..204..131K,
       author = {{Kasper}, Justin C. and {Abiad}, Robert and {Austin}, Gerry and {Balat-Pichelin}, Marianne and {Bale}, Stuart D. and {Belcher}, John W. and {Berg}, Peter and {Bergner}, Henry and {Berthomier}, Matthieu and {Bookbinder}, Jay and {Brodu}, Etienne and {Caldwell}, David and {Case}, Anthony W. and {Chandran}, Benjamin D.~G. and {Cheimets}, Peter and {Cirtain}, Jonathan W. and {Cranmer}, Steven R. and {Curtis}, David W. and {Daigneau}, Peter and {Dalton}, Greg and {Dasgupta}, Brahmananda and {DeTomaso}, David and {Diaz-Aguado}, Millan and {Djordjevic}, Blagoje and {Donaskowski}, Bill and {Effinger}, Michael and {Florinski}, Vladimir and {Fox}, Nichola and {Freeman}, Mark and {Gallagher}, Dennis and {Gary}, S. Peter and {Gauron}, Tom and {Gates}, Richard and {Goldstein}, Melvin and {Golub}, Leon and {Gordon}, Dorothy A. and {Gurnee}, Reid and {Guth}, Giora and {Halekas}, Jasper and {Hatch}, Ken and {Heerikuisen}, Jacob and {Ho}, George and {Hu}, Qiang and {Johnson}, Greg and {Jordan}, Steven P. and {Korreck}, Kelly E. and {Larson}, Davin and {Lazarus}, Alan J. and {Li}, Gang and {Livi}, Roberto and {Ludlam}, Michael and {Maksimovic}, Milan and {McFadden}, James P. and {Marchant}, William and {Maruca}, Bennet A. and {McComas}, David J. and {Messina}, Luciana and {Mercer}, Tony and {Park}, Sang and {Peddie}, Andrew M. and {Pogorelov}, Nikolai and {Reinhart}, Matthew J. and {Richardson}, John D. and {Robinson}, Miles and {Rosen}, Irene and {Skoug}, Ruth M. and {Slagle}, Amanda and {Steinberg}, John T. and {Stevens}, Michael L. and {Szabo}, Adam and {Taylor}, Ellen R. and {Tiu}, Chris and {Turin}, Paul and {Velli}, Marco and {Webb}, Gary and {Whittlesey}, Phyllis and {Wright}, Ken and {Wu}, S.~T. and {Zank}, Gary},
        title = "{Solar Wind Electrons Alphas and Protons (SWEAP) Investigation: Design of the Solar Wind and Coronal Plasma Instrument Suite for Solar Probe Plus}",
      journal = {\ssr},
         year = 2016,
        month = dec,
       volume = {204},
       number = {1-4},
        pages = {131-186},
          doi = {10.1007/s11214-015-0206-3},
       adsurl = {https://ui.adsabs.harvard.edu/abs/2016SSRv..204..131K}
}

@article{fox2016solar,
  title={The solar probe plus mission: humanity’s first visit to our star},
  author={Fox, NJ and Velli, MC and Bale, SD and Decker, R and Driesman, A and Howard, RA and Kasper, Justin C and Kinnison, J and Kusterer, M and Lario, D and others},
  journal={Space Science Reviews},
  volume={204},
  number={1},
  pages={7--48},
  year={2016},
  publisher={Springer}
}

@article{de2020switchbacks,
  title={Switchbacks in the near-Sun magnetic field: long memory and impact on the turbulence cascade},
  author={de Wit, Thierry Dudok and Krasnoselskikh, Vladimir V and Bale, Stuart D and Bonnell, John W and Bowen, Trevor A and Chen, Christopher HK and Froment, Clara and Goetz, Keith and Harvey, Peter R and Jagarlamudi, Vamsee Krishna and others},
  journal={The Astrophysical Journal Supplement Series},
  volume={246},
  number={2},
  pages={39},
  year={2020},
  publisher={IOP Publishing}
}

@article{pecora2022magnetic,
  title={Magnetic switchback occurrence rates in the inner heliosphere: parker solar probe and 1 au},
  author={Pecora, Francesco and Matthaeus, William H and Primavera, Leonardo and Greco, Antonella and Chhiber, Rohit and Bandyopadhyay, Riddhi and Servidio, Sergio},
  journal={The Astrophysical Journal Letters},
  volume={929},
  number={1},
  pages={L10},
  year={2022},
  publisher={IOP Publishing}
}

@article{kasper2019alfvenic,
  title={Alfv{\'e}nic velocity spikes and rotational flows in the near-Sun solar wind},
  author={Kasper, Justin C and Bale, Stuart D and Belcher, John Winston and Berthomier, Matthieu and Case, Anthony W and Chandran, Benjamin DG and Curtis, DW and Gallagher, D and Gary, SP and Golub, L and others},
  journal={Nature},
  volume={576},
  number={7786},
  pages={228--231},
  year={2019},
  publisher={Nature Publishing Group UK London}
}

@article{bale2019highly,
  title={Highly structured slow solar wind emerging from an equatorial coronal hole},
  author={Bale, SD and Badman, ST and Bonnell, JW and Bowen, TA and Burgess, D and Case, AW and Cattell, CA and Chandran, BDG and Chaston, CC and Chen, CHK and others},
  journal={Nature},
  volume={576},
  number={7786},
  pages={237--242},
  year={2019},
  publisher={Nature Publishing Group UK London}
}

@article{phan2020parker,
  title={Parker solar probe in situ observations of magnetic reconnection exhausts during encounter 1},
  author={Phan, TD and Bale, SD and Eastwood, JP and Lavraud, B and Drake, JF and Oieroset, M and Shay, MA and Pulupa, M and Stevens, M and MacDowall, RJ and others},
  journal={The Astrophysical Journal Supplement Series},
  volume={246},
  number={2},
  pages={34},
  year={2020},
  publisher={IOP Publishing}
}

@article{tenerani2021evolution,
  title={Evolution of switchbacks in the inner heliosphere},
  author={Tenerani, Anna and Sioulas, Nikos and Matteini, Lorenzo and Panasenco, Olga and Shi, Chen and Velli, Marco},
  journal={The Astrophysical Journal Letters},
  volume={919},
  number={2},
  pages={L31},
  year={2021},
  publisher={IOP Publishing}
}

@article{mozer2021origin,
  title={On the origin of switchbacks observed in the solar wind},
  author={Mozer, Forrest S and Bale, SD and Bonnell, JW and Drake, JF and Hanson, ELM and Mozer, Michael C},
  journal={The Astrophysical Journal},
  volume={919},
  number={1},
  pages={60},
  year={2021},
  publisher={IOP Publishing}
}

@article{sioulas2025propagation,
  title={On the Propagation and Damping of Alfvenic Fluctuations in the Outer Solar Corona and Solar Wind},
  author={Sioulas, Nikos and Velli, Marco and Shi, Chen and Bowen, Trevor A and Mallet, Alfred and Verdini, Andrea and Chandran, BDG and Tenerani, Anna and Dakeyo, Jean-Baptiste and Bale, Stuart D and others},
  journal={arXiv preprint arXiv:2510.10106},
  year={2025}
}

@article{livi2022solar,
  title={The solar probe analyzer—ions on the Parker Solar Probe},
  author={Livi, Roberto and Larson, Davin E and Kasper, Justin C and Abiad, Robert and Case, Anthony W and Klein, Kristopher G and Curtis, David W and Dalton, Gregory and Stevens, Michael and Korreck, Kelly E and others},
  journal={The Astrophysical Journal},
  volume={938},
  number={2},
  pages={138},
  year={2022},
  publisher={The American Astronomical Society}
}

@article{sioulas2026generation,
  title={Generation and Expansion-Driven Growth of Switchbacks in the Outer Solar Corona and Solar Wind},
  author={Sioulas, Nikos and Velli, Marco and Shi, Chen and Matteini, Lorenzo and Bowen, Trevor A and Mallet, Alfred and Larosa, A and Tenerani, Anna and Horbury, Timothy S},
  journal={arXiv preprint arXiv:2602.03724},
  year={2026}
}

@article{meyrand2025reflection,
  title={Reflection-driven turbulence in the super-Alfv{\'e}nic solar wind},
  author={Meyrand, Romain and Squire, Jonathan and Mallet, Alfred and Chandran, Benjamin DG},
  journal={Journal of plasma physics},
  volume={91},
  number={1},
  pages={E29},
  year={2025},
  publisher={Cambridge University Press}
}

@ARTICLE{2021ApJ...923..174B,
       author = {{Bale}, S.~D. and {Horbury}, T.~S. and {Velli}, M. and {Desai}, M.~I. and {Halekas}, J.~S. and {McManus}, M.~D. and {Panasenco}, O. and {Badman}, S.~T. and {Bowen}, T.~A. and {Chandran}, B.~D.~G. and {Drake}, J.~F. and {Kasper}, J.~C. and {Laker}, R. and {Mallet}, A. and {Matteini}, L. and {Phan}, T.~D. and {Raouafi}, N.~E. and {Squire}, J. and {Woodham}, L.~D. and {Woolley}, T.},
        title = "{A Solar Source of Alfv{\'e}nic Magnetic Field Switchbacks: In Situ Remnants of Magnetic Funnels on Supergranulation Scales}",
      journal = {\apj},
         year = 2021,
        month = dec,
       volume = {923},
       number = {2},
          eid = {174},
        pages = {174},
          doi = {10.3847/1538-4357/ac2d8c},
archivePrefix = {arXiv},
       eprint = {2109.01069},
 primaryClass = {astro-ph.SR},
       adsurl = {https://ui.adsabs.harvard.edu/abs/2021ApJ...923..174B}
}

@ARTICLE{2022ApJ...934..152S,
       author = {{Shi}, Chen and {Panasenco}, Olga and {Velli}, Marco and {Tenerani}, Anna and {Verniero}, Jaye L. and {Sioulas}, Nikos and {Huang}, Zesen and {Brosius}, A. and {Bale}, Stuart D. and {Klein}, Kristopher and {Kasper}, Justin and {de Wit}, Thierry Dudok and {Goetz}, Keith and {Harvey}, Peter R. and {MacDowall}, Robert J. and {Malaspina}, David M. and {Pulupa}, Marc and {Larson}, Davin and {Livi}, Roberto and {Case}, Anthony and {Stevens}, Michael},
        title = "{Patches of Magnetic Switchbacks and Their Origins}",
      journal = {\apj},
         year = 2022,
        month = aug,
       volume = {934},
       number = {2},
          eid = {152},
        pages = {152},
          doi = {10.3847/1538-4357/ac7c11},
archivePrefix = {arXiv},
       eprint = {2206.03807},
 primaryClass = {astro-ph.SR},
       adsurl = {https://ui.adsabs.harvard.edu/abs/2022ApJ...934..152S}
}

@article{horbury2020solar,
  title={The solar orbiter magnetometer},
  author={Horbury, TS and O’brien, H and Carrasco Blazquez, I and Bendyk, M and Brown, P and Hudson, R and Evans, V and Oddy, TM and Carr, CM and Beek, TJ and others},
  journal={Astronomy \& Astrophysics},
  volume={642},
  pages={A9},
  year={2020},
  publisher={edp sciences}
}

@article{lepping1995wind,
  title={The WIND magnetic field investigation},
  author={Lepping, RP and Ac{\~u}na, MH and Burlaga, LF and Farrell, WM and Slavin, JA and Schatten, KH and Mariani, F and Ness, NF and Neubauer, FM and Whang, YC and others},
  journal={Space Science Reviews},
  volume={71},
  number={1},
  pages={207--229},
  year={1995},
  publisher={Springer}
}

@article{smith1998ace,
  title={The ACE magnetic fields experiment},
  author={Smith, Charles W and L'Heureux, Jacques and Ness, Norman F and Acuna, Mario H and Burlaga, Leonard F and Scheifele, John},
  journal={Space Science Reviews},
  volume={86},
  number={1},
  pages={613--632},
  year={1998},
  publisher={Springer}
}

@article{balogh1992magnetic,
  title={The magnetic field investigation on the ULYSSES mission-Instrumentation and preliminary scientific results},
  author={Balogh, A and Beek, To J and Forsyth, RJ and Hedgecock, PC and Marquedant, RJ and Smith, EJ and Southwood, DJ and Tsurutani, BT},
  journal={Astronomy and Astrophysics Supplement Series (ISSN 0365-0138), vol. 92, no. 2, Jan. 1992, p. 221-236. Research supported by SERC.},
  volume={92},
  pages={221--236},
  year={1992}
}

@article{larosa2021switchbacks,
  title={Switchbacks: statistical properties and deviations from Alfv{\'e}nicity},
  author={Larosa, A and Krasnoselskikh, V and Dudok de Wit, T and Agapitov, O and Froment, C and Jagarlamudi, VK and Velli, M and Bale, SD and Case, AW and Goetz, K and others},
  journal={Astronomy \& Astrophysics},
  volume={650},
  pages={A3},
  year={2021},
  publisher={EDP Sciences}
}

@article{hernandez2021impact,
  title={Impact of switchbacks on turbulent cascade and energy transfer rate in the inner heliosphere},
  author={Hern{\'a}ndez, Carlos S and Sorriso-Valvo, Luca and Bandyopadhyay, Riddhi and Chasapis, Alexandros and V{\'a}sconez, Christian L and Marino, Raffaele and Pezzi, Oreste},
  journal={The Astrophysical Journal Letters},
  volume={922},
  number={1},
  pages={L11},
  year={2021},
  publisher={The American Astronomical Society}
}

@article{rivera2024situ,
  title={In situ observations of large-amplitude Alfv{\'e}n waves heating and accelerating the solar wind},
  author={Rivera, Yeimy J and Badman, Samuel T and Stevens, Michael L and Verniero, Jaye L and Stawarz, Julia E and Shi, Chen and Raines, Jim M and Paulson, Kristoff W and Owen, Christopher J and Niembro, Tatiana and others},
  journal={Science},
  volume={385},
  number={6712},
  pages={962--966},
  year={2024},
  publisher={American Association for the Advancement of Science}
}

@article{wyper2026magnetic,
  title={Magnetic switchback formation: a review of proposed mechanisms},
  author={Wyper, Peter F and Squire, Jonathan and Pariat, Etienne and Agapitov, Oleksiy V and Drake, Jim F and Magyar, Norbert and Matthaeus, William H and Matteini, Lorenzo and Ruffolo, David and R{\'e}ville, Victor and others},
  journal={Space Science Reviews},
  volume={222},
  number={4},
  pages={43},
  year={2026},
  publisher={Springer}
}

@article{tziotziou2023vortex,
  title={Vortex motions in the solar atmosphere: definitions, theory, observations, and modelling},
  author={Tziotziou, K and Scullion, E and Shelyag, Sergiy and Steiner, O and Khomenko, E and Tsiropoula, G and Canivete Cuissa, JR and Wedemeyer, S and Kontogiannis, I and Yadav, Nitin and others},
  journal={Space Science Reviews},
  volume={219},
  number={1},
  pages={1},
  year={2023},
  publisher={Springer}
}

@article{zank2020origin,
  title={The origin of switchbacks in the solar corona: linear theory},
  author={Zank, GP and Nakanotani, M and Zhao, L-L and Adhikari, L and Kasper, J},
  journal={The Astrophysical Journal},
  volume={903},
  number={1},
  pages={1},
  year={2020},
  publisher={The American Astronomical Society}
}

@article{touresse2024propagation,
  title={Propagation of untwisting solar jets from the low-beta corona into the super-Alfv{\'e}nic wind: Testing a solar origin scenario for switchbacks},
  author={Touresse, Jade and Pariat, Etienne and Froment, Clara and Aslanyan, Valentin and Wyper, Peter F and Seyfritz, Louis},
  journal={Astronomy \& Astrophysics},
  volume={692},
  pages={A71},
  year={2024},
  publisher={EDP Sciences}
}

@article{toth2023theory,
  title={Theory of magnetic switchbacks fully supported by Parker Solar Probe observations},
  author={Toth, Gabor and Velli, Marco and van der Holst, Bart},
  journal={The Astrophysical Journal},
  volume={957},
  number={2},
  pages={95},
  year={2023},
  publisher={The American Astronomical Society}
}

@article{schwadron2021switchbacks,
  title={Switchbacks explained: super-Parker fields—the other side of the sub-Parker spiral},
  author={Schwadron, NA and McComas, DJ},
  journal={The Astrophysical Journal},
  volume={909},
  number={1},
  pages={95},
  year={2021},
  publisher={The American Astronomical Society}
}

@article{ruffolo2020shear,
  title={Shear-driven transition to isotropically turbulent solar wind outside the Alfv{\'e}n critical zone},
  author={Ruffolo, D and Matthaeus, William H and Chhiber, Rohit and Usmanov, Arcadi V and Yang, Yan and Bandyopadhyay, Riddhi and Parashar, TN and Goldstein, Melvyn L and DeForest, CE and Wan, Minping and others},
  journal={The Astrophysical Journal},
  volume={902},
  number={2},
  pages={94},
  year={2020},
  publisher={The American Astronomical Society}
}

@article{huang2023structure,
  title={The structure and origin of switchbacks: Parker Solar Probe observations},
  author={Huang, Jia and Kasper, JC and Fisk, LA and Larson, Davin E and McManus, Michael D and Chen, CHK and Martinovi{\'c}, Mihailo M and Klein, KG and Thomas, Luke and Liu, Mingzhe and others},
  journal={The Astrophysical Journal},
  volume={952},
  number={1},
  pages={33},
  year={2023},
  publisher={The American Astronomical Society}
}

@article{badman2026properties,
  title={Properties of Magnetic Switchbacks in the Near-Sun Solar Wind},
  author={Badman, Samuel T and Fargette, Na{\"\i}s and Matteini, Lorenzo and Agapitov, Oleksiy V and Akhavan-Tafti, Mojtaba and Bale, Stuart D and Bharati Das, Srijan and Bizien, Nina and Bowen, Trevor A and Dudok de Wit, Thierry and others},
  journal={Space Science Reviews},
  volume={222},
  number={1},
  pages={14},
  year={2026},
  publisher={Springer}
}

@article{wu2021large,
  title={Large amplitude switchback turbulence: possible magnetic velocity alignment structures},
  author={Wu, Honghong and Tu, Chuanyi and Wang, Xin and Yang, Liping},
  journal={The Astrophysical Journal},
  volume={911},
  number={2},
  pages={73},
  year={2021},
  publisher={The American Astronomical Society}
}

@article{agapitov2023constraints,
  title={Constraints on the Alfv{\'e}nicity of switchbacks},
  author={Agapitov, OV and Drake, JF and Swisdak, Marc and Choi, K-E and Raouafi, Nour},
  journal={The Astrophysical Journal Letters},
  volume={959},
  number={2},
  pages={L21},
  year={2023},
  publisher={The American Astronomical Society}
}

@article{jagarlamudi2023occurrence,
  title={Occurrence and evolution of switchbacks in the inner heliosphere: Parker Solar Probe observations},
  author={Jagarlamudi, Vamsee Krishna and Raouafi, NE and Bourouaine, S and Mostafavi, P and Larosa, A and Perez, JC},
  journal={The Astrophysical Journal Letters},
  volume={950},
  number={1},
  pages={L7},
  year={2023},
  publisher={The American Astronomical Society}
}

@article{meng2022analysis,
  title={Analysis of the distribution, rotation and scale characteristics of solar wind switchbacks: comparison between the first and second encounters of Parker Solar Probe},
  author={Meng, Ming-Ming and Liu, Ying D and Chen, Chong and Wang, Rui},
  journal={Research in Astronomy and Astrophysics},
  volume={22},
  number={3},
  pages={035018},
  year={2022},
  publisher={National Astromonical Observatories, CAS and IOP Publishing}
}

@article{huang2023parker,
  title={Parker Solar Probe observations of high plasma $\beta$ solar wind from the streamer belt},
  author={Huang, Jia and Kasper, Justin Christophe and Larson, Davin E and McManus, Michael D and Whittlesey, P and Livi, Roberto and Rahmati, Ali and Romeo, Orlando and Klein, KG and Sun, Weijie and others},
  journal={The Astrophysical Journal Supplement Series},
  volume={265},
  number={2},
  pages={47},
  year={2023},
  publisher={The American Astronomical Society}
}

@article{halekas2023quantifying,
  title={Quantifying the energy budget in the solar wind from 13.3 to 100 solar radii},
  author={Halekas, JS and Bale, SD and Berthomier, M and Chandran, BDG and Drake, JF and Kasper, JC and Klein, KG and Larson, DE and Livi, R and Pulupa, MP and others},
  journal={The Astrophysical Journal},
  volume={952},
  number={1},
  pages={26},
  year={2023},
  publisher={The American Astronomical Society}
}

@article{fargette2021characteristic,
  title={Characteristic scales of magnetic switchback patches near the sun and their possible association with solar supergranulation and granulation},
  author={Fargette, Na{\"\i}s and Lavraud, Benoit and Rouillard, Alexis P and R{\'e}ville, Victor and Dudok De Wit, Thierry and Froment, Clara and Halekas, Jasper S and Phan, Tai D and Malaspina, David M and Bale, Stuart D and others},
  journal={The Astrophysical Journal},
  volume={919},
  number={2},
  pages={96},
  year={2021},
  publisher={The American Astronomical Society}
}

@article{payne2026evolution,
  title={Evolution of Magnetic Deflections at a Conversion Layer near the Alfv{\'e}n Surface},
  author={Payne, Dominic and Akhavan-Tafti, Mojtaba and Goodwill, Joshua and Badman, Samuel and Bandyopadhyay, Riddhi and Zank, Gary and Adhikari, Subash and Matthaeus, William and Shi, Chen and Stevens, Michael and others},
  journal={The Astrophysical Journal Letters},
  volume={1001},
  number={2},
  pages={L29},
  year={2026},
  publisher={The American Astronomical Society}
}

@article{shi2021alfvenic,
  title={Alfv{\'e}nic versus non-Alfv{\'e}nic turbulence in the inner heliosphere as observed by Parker Solar Probe},
  author={Shi, Chen and Velli, Marco and Panasenco, Olga and Tenerani, Anna and R{\'e}ville, Victor and Bale, Stuart D and Kasper, Justin and Korreck, Kelly and Bonnell, JW and de Wit, T Dudok and others},
  journal={Astronomy \& Astrophysics},
  volume={650},
  pages={A21},
  year={2021},
  publisher={EDP Sciences}
}

@article{mallet2026evolution,
  title={Evolution and impact of switchbacks throughout the heliosphere},
  author={Mallet, Alfred and Shi, Chen and Tenerani, Anna and Agapitov, Oleksiy and Akhavan-Tafti, Mojtaba and Badman, Samuel and Bizien, Nina and Bowen, Trevor and Desai, Mihir I and Drake, JF and others},
  journal={arXiv preprint arXiv:2607.02709},
  year={2026}
}
\bibliographystyle{aasjournalv7}



\end{document}